\documentclass[a4paper,twoside,twocolumn,english,superscriptaddress,aps,showpacs]{revtex4}
\usepackage[T1]{fontenc}
\usepackage[latin9]{inputenc}
\usepackage{amsmath}
\usepackage{graphicx}
\usepackage{amssymb}

\makeatletter

\newcommand{\lyxdot}{.}

\@ifundefined{textcolor}{}
{%
 \definecolor{BLACK}{gray}{0}
 \definecolor{WHITE}{gray}{1}
 \definecolor{RED}{rgb}{1,0,0}
 \definecolor{GREEN}{rgb}{0,1,0}
 \definecolor{BLUE}{rgb}{0,0,1}
 \definecolor{CYAN}{cmyk}{1,0,0,0}
 \definecolor{MAGENTA}{cmyk}{0,1,0,0}
 \definecolor{YELLOW}{cmyk}{0,0,1,0}
 }

\makeatother

\usepackage{babel}

\begin{document}

\title{Ground states of dipolar gases in quasi-1D ring traps}

\author{Sascha Zöllner}

\thanks{Present address: Helmholtz Center Munich, 85764 Neuherberg, Germany}

\email{zoellner@nbi.dk}

\affiliation{The Niels Bohr International Academy, The Niels Bohr Institute, Blegdamsvej
17, 2100 Copenhagen, Denmark}


\date{December 14, 2011}
\begin{abstract}
We compute the ground state of dipoles in a quasi-one-dimensional
ring trap using few-body techniques combined with analytic arguments.\emph{
}The effective interaction between two dipoles depends on their center-of-mass
coordinate and can be tuned by varying the angle between dipoles and
the plane of the ring. For weak enough interactions, the state resembles
a weakly interacting Fermi gas or an (inhomogeneous) Lieb-Liniger
gas. A mapping between the Lieb-Liniger and the dipolar-gas parameters
in and beyond the Born approximation is established, and we discuss
the effect of inhomogeneities based on a local-density approximation.
For strongly repulsive interactions, the system exhibits crystal-like
localization of the particles. Their inhomogeneous distribution may
be understood in terms of a simple few-body model as well as a local-density
approximation. In the case of partially attractive interactions, clustered
states form for strong enough coupling, and the dependence of the
state on particle number and orientation angle of the dipoles is discussed
analytically.
\end{abstract}
\maketitle

\section{Introduction}

The creation and study of so-called dipolar quantum gases have recently
become a major research focus \cite{baranov08,lahaye09}. These gases,
such as ultracold atoms with magnetic dipole moments (e.g., Cr \cite{lahaye07}
or Dy \cite{lu10}) or polar molecules (such as KRb \cite{ospelkaus10},
LiCs \cite{deiglmayr08} or RbCs \cite{sage05}), are dominated by
dipole-dipole rather than van-der-Waals interactions.\textbf{ }Combining
a high level of control with long-ranged and strongly anisotropic\textbf{
}interactions, dipolar gases offer both the simulation of elusive
quantum states in condensed-matter or nuclear physics and the design
of exotic novel quantum phases. 

The question naturally arises how a dipolar gas behaves when confined
to lower dimensions. This is important experimentally, because the
lower dimensionality may help alleviate the collisional instability
toward head-to-tail alignment of the dipoles \cite{koch08,ni10},
and conceptually, given that the enhanced quantum fluctuations in
lower dimensions give rise to intriguing physics. 

In particular, in strictly one dimension (1D), dipolar gases have
been shown to exhibit Luttinger-liquid behavior (see, e.g., \cite{arkhipov05,citro07,depalo08,pedri08}).
A finite, quasi-1D transverse confinement may both remove the short-distance
divergence of the dipolar interactions \cite{sinha07} and alter the
short-range s-wave interactions, which may lead to intriguing physics
like a roton instability toward a density wave. In spite of displaying
long-range interactions, this system still is an effectively homogeneous
1D system, so long as the dipoles have a common orientation. In a
linear geometry, the anisotropy enters only once the system ceases
to be one-dimensional, in which case {}``zig-zag'' chains or other
higher-dimensional configurations may show up \cite{astrakharchik08a}.

A dipolar system exhibiting truly 1D physics as well as anisotropic
interactions can be achieved in curved lower-dimensional geometries.
Nontrivial geometries are interesting not only in the context of dipolar
interactions \cite{dutta06,law08,abad10,abad11,maik11}, but also
for studying, e.g., persistent currents \cite{ramanathan11,amico05}.
In the simplest case of a quasi-1D ring trap, which can be realized
in the context of cold atoms \cite{gupta05,arnold06,heathcote08,henderson09},
the effective 1D dipole interaction between two particles becomes
inhomogeneous in the sense that it acquires a dependence on their
center-of-mass coordinate. In this paper, we show how by varying both
the overall strength as well as the degree of inhomogeneity, interesting
regimes are found in such a system -- such as a 1D Bose-gas-like phase
with center-of-mass dependent interaction $\gamma$, a Wigner crystal
with inhomogeneous lattice spacing, and self-bound clusters of identical
fermions. These are explained on the basis of numerical few-body calculations
as well as by deriving simple analytic models. A short account of
these findings has been published recently \cite{zoellner10a}.

This paper is organized as follows. Section \ref{sec:theory} introduces
the effective Hamiltonian and discusses aspects of the corresponding
two-body problem. In Sec.~\ref{sec:Homogeneous}, we investigate
the many-body ground-state in the limit of homogeneous (repulsive)
interactions, i.e., dipoles aligned perpendicular to the plane of
the ring. Section~\ref{sec:Repulsion} focuses on the case where
the interaction is purely repulsive but inhomogeneous. Partially attractive
interactions, as occur for small enough angles between dipoles and
the ring plane, are studied in Sec.~\ref{sec:Attraction}. Appendix~\ref{sec:method}
contains a concise introduction to the numerically exact multi-configurational
time-dependent Hartree method employed for the computation of few-body
ground-state properties.\newpage{}

\section{Model \label{sec:theory}}

\subsection{Hamiltonian}

\begin{figure}
\begin{centering}
\includegraphics[width=0.6\columnwidth]{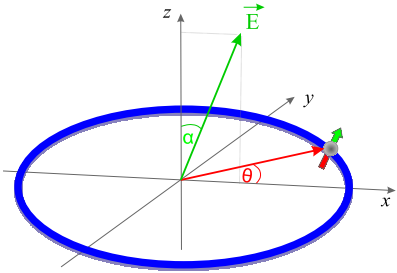}
\par\end{centering}

\caption{(color online) Sketch of the ring-shaped trap (radius $R$, width
$a_{\perp}\ll R$) in the $xy$ plane, with the dipoles moments $\mathbf{d}=d\left(\sin\alpha,0,\cos\alpha\right)^{\top}$
polarized in the $xz$ direction by an external field $\mathbf{E}$.
\label{fig:setup}}
\end{figure}

We consider a system of $N$ identical particles (bosons or fermions)
of mass $m$, confined in a ring-shaped trapping potential. The ring
(radius $R$) is taken to lie in the $xy$-plane (Fig.\ \ref{fig:setup}).
The particles (e.g., atoms with a magnetic moment or polar molecules)
have a dipole moment $\mathbf{d}=d\left(\sin\alpha,0,\cos\alpha\right)$
aligned in the $xz$-plane by an external field, at an angle $\alpha$
to the $z$-axis. The interaction between two dipoles in free space
is $V(\mathbf{r})=D^{2}(1-3\cos^{2}\theta_{rd})/r^{3}$, where $\mathbf{r}$
is the separation between the dipoles and $\theta_{rd}$ is the angle
between $\mathbf{r}$ and $\mathbf{d}$; furthermore $D^{2}=d^{2}/4\pi\epsilon_{0}$
for electric dipoles and $d^{2}\mu_{0}/4\pi$ for magnetic ones. Note
that for small separations, the far-field dipolar interaction is no
longer valid, and short-range interactions dominate. In order to map
out the genuine dipolar physics, we assume that short-ranged forces
can be omitted on the length scale of interest. One might imagine
tuning them to zero using optical or magnetic Feshbach management;
however, it may be important to have short-range repulsion in order
to stabilize the dipoles.

We now focus on the limit of a ring potential with tight harmonic
confinement in the transverse direction. Then, for sufficiently weak
interaction strength, the transverse motion is frozen in the lowest-energy
mode, which is a Gaussian $\varphi_{_{0}}(\rho)=e^{-\rho^{2}/2a_{\perp}^{2}}/\sqrt{\pi}a_{\perp}$
of spatial extent $a_{\perp}\equiv$$\sqrt{\hbar/m\omega_{\perp}}\ll R$
\cite{sinha07}. Averaging out the transverse degrees of freedom over
the reduced density matrix $\hat{\rho}_{\perp}=|\varphi_{0}\rangle\langle\varphi_{_{0}}|$
, one arrives at an effective 1D Hamiltonian \cite{dutta06}\begin{equation}
H=-\frac{\hbar^{2}}{2mR^{2}}\sum_{i=1}^{N}\frac{\partial^{2}}{\partial\theta_{i}^{2}}+\sum_{i<j}V_{{\rm 1D}}(\theta_{i},\theta_{j}),\label{eq:Hamiltonian}\end{equation}
where the angle $\theta_{i}$ specifies the position of particle $i$
on the ring. For $R\gg a_{\perp}$, the effective interaction takes
the form $V_{{\rm 1D}}(\theta_{1},\theta_{2})=V_{{\rm CM}}\left(\Theta\right)V_{{\rm rel}}(\vartheta)$,
where $\Theta=(\theta_{1}+\theta_{2})/2$ is the center-of-mass (CM)
angle of the two dipoles and $\vartheta=\theta_{1}-\theta_{2}$ is
the relative angle. In terms of $s=2R|\sin(\vartheta/2)|/a_{\perp}$,
the dependence on the relative angle is given by \begin{equation}
V_{{\rm rel}}(\vartheta)=\sqrt{2\pi}(1+s^{2})e^{s^{2}/2}\mathrm{erfc}(s/\sqrt{2})-2s;\label{eq:Vrel}\end{equation}
 moreover, \[
V_{{\rm CM}}(\Theta)=\frac{D^{2}}{4a_{\perp}^{3}}(1-3\sin^{2}\alpha\sin^{2}\Theta).\]

\begin{figure}
\begin{centering}
\includegraphics[width=0.8\columnwidth]{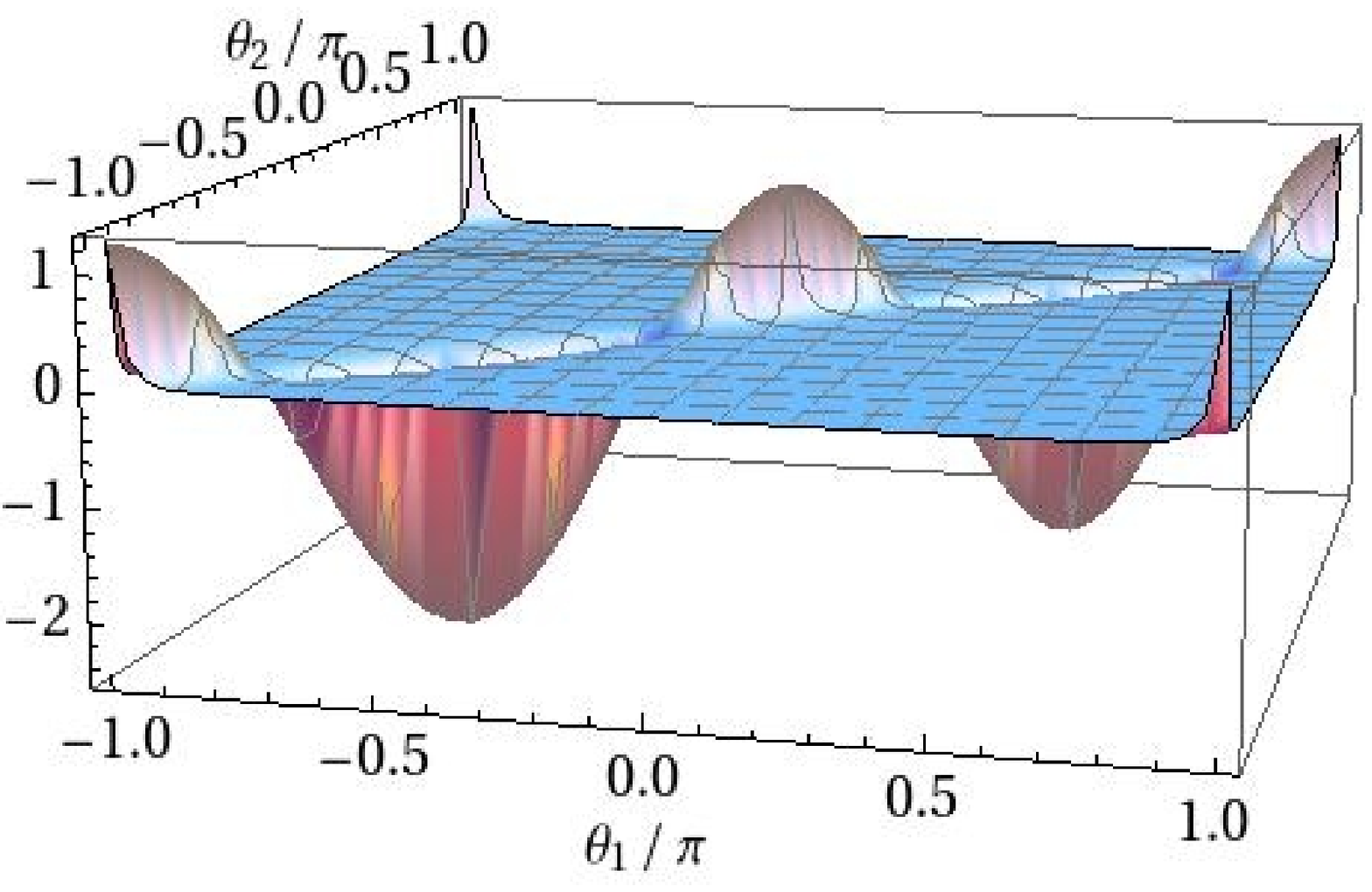}
\par\end{centering}

\begin{centering}
\includegraphics[width=0.45\columnwidth]{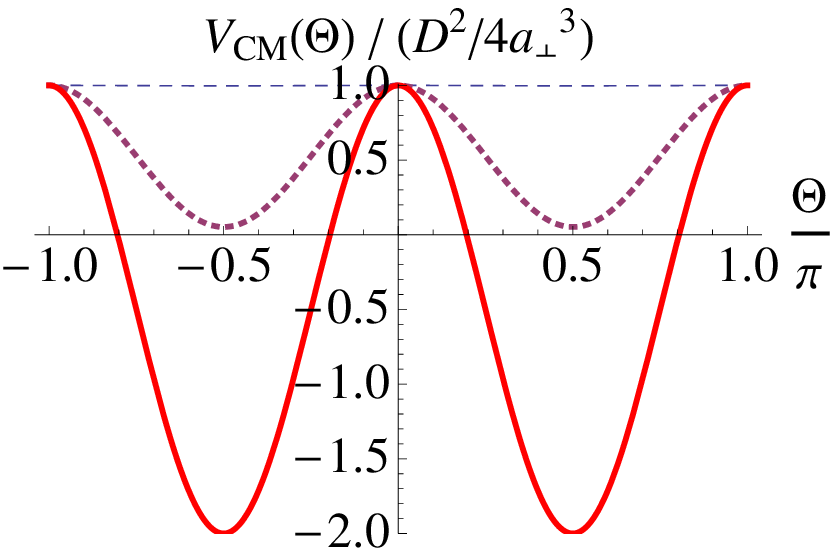}\includegraphics[width=0.45\columnwidth]{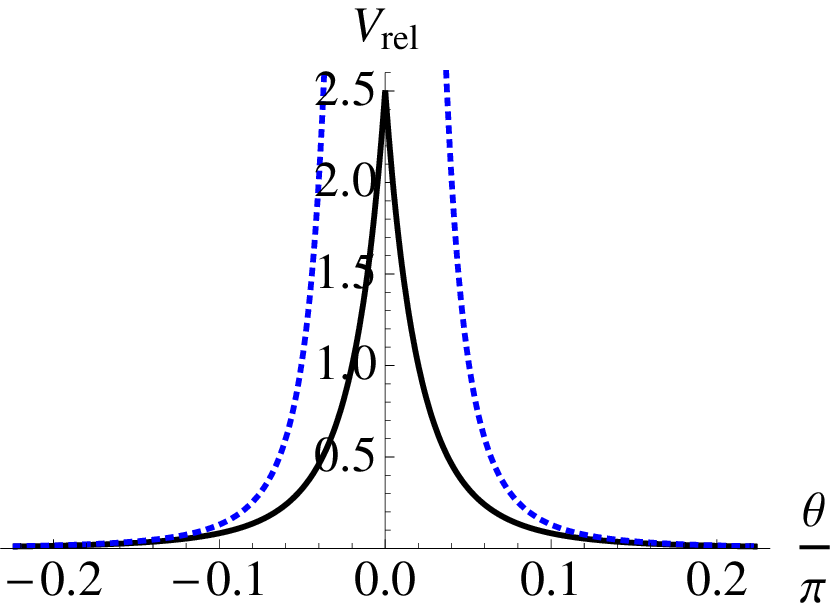}
\par\end{centering}

\caption{(color online) Effective interaction: (a) Potential-energy surface
$V_{1\mathrm{D}}(\theta_{1},\theta_{2})$ (here for $\alpha=\frac{\pi}{2}$).
(b) Center-of-mass-dependent interaction strength $V_{\mathrm{CM}}(\Theta)$
for $\alpha=0,\,0.19\pi$ and $\frac{\pi}{2}$. (c) Relative-coordinate
potential $V_{\mathrm{rel}}(\vartheta)$ {[}thick line{]} for $R/a_{\perp}=10$,
including the long-range asymptotics (thin line). \label{fig:potential}}
\end{figure}

Note that the effective interaction $V_{\mathrm{1D}}$ is \emph{inhomogeneous}
in the angular coordinates: This reflects the anisotropy of the underlying
3D interaction, which favors configurations where the two dipoles
are arranged preferably parallel to $\mathbf{d}$, i.e., such that
$\cos\theta_{rd}=\sin\alpha\sin\left({\scriptstyle \frac{\theta_{1}+\theta_{2}}{2}}\right)$
is maximized, with the particles restricted to the ring. Thus the
CM potential, shown in Fig.~\ref{fig:potential}, has minima at $\Theta=\pm\frac{\pi}{2}$,
which become more pronounced for larger $\alpha$. For \[
\alpha>\alpha_{\mathrm{c}}\equiv\arcsin\left(\frac{1}{\sqrt{3}}\right)\approx0.196\pi,\]
 the potential acquires attractive regions. The potential should be
viewed as a potential-energy surface which depends on both particles'
coordinates rather than just the distance between them, as sketched
in Fig.~\ref{fig:potential}.

The derivation of the effective interaction (\ref{eq:Vrel}) assumes
that the single-mode approximation  is valid, i.e., that only the
lowest transverse mode $\varphi_{0}$ is occupied. This is equivalent
to first-order perturbation theory or, in the language of scattering
theory, the Born approximation (where $\varphi_{0}(\rho)e^{iqx}$
replaces the 3D plane-wave states in the presence of confinement).
This is valid so long as the interaction energy is small compared
with the transverse level spacing, $\hbar\omega_{\perp}$, i.e., if
$r_{d}\ll a_{\perp}$. In that case, second-order perturbation theory
predicts negative corrections $O(\frac{r_{d}}{a_{\perp}})^{2}$ due
to virtual excitation of higher transverse levels. For interaction
energies of order $\hbar\omega_{\perp}$, confinement-induced resonances
may be expected, similar to those for contact interactions under transverse
confinement \cite{Olshanii1998a,sinha07}.

\subsection{Two-body problem: Relative motion \label{sub:Relative-problem}}

Let us briefly discuss some properties of the two-body relative problem,
given by the Hamiltonian \[
h_{\mathrm{rel}}=-\frac{\hbar^{2}}{mR^{2}}\frac{\partial^{2}}{\partial\vartheta^{2}}+\frac{D^{2}}{4a_{\perp}^{3}}V_{\mathrm{rel}}\left(\frac{r}{a_{\perp}}\right),\quad r\equiv2R\sin\frac{\vartheta}{2}.\]
Strictly speaking, this is relevant only if CM and relative motions
decouple -- as in the case of dipoles perpendicular to the plane of
the ring ($\alpha=0$), but also, more generally, for a linear geometry
\cite{sinha07}. However, it will also provide useful insight in cases
where the particle spacing is small compared with the length scale
on which $V_{CM}$ varies, so that one may locally replace $D^{2}/4a_{\perp}^{3}\equiv V_{\mathrm{CM}}(\Theta)$
at fixed $\Theta$.

For distances $|r|\gg a_{\perp}$, the free-space dipolar potential
is restored, so that the Hamiltonian becomes $a_{\perp}$ independent,
\[
\frac{D^{2}}{4a_{\perp}^{3}}V_{\mathrm{rel}}\left(\frac{r}{a_{\perp}}\right)\stackrel{|r|\gg a_{\perp}}{\simeq}\frac{D^{2}}{r^{3}}.\]
Its strength is characterized by the \emph{dipolar length }\[
r_{d}\equiv\frac{2m}{\hbar^{2}}D^{2}.\]
 At short distances, the transverse average introduces a short-range
{}``cutoff'' $a_{\perp}$. This attenuates the $r=0$ divergence
into a peak of order $D^{2}/a_{\perp}^{3}$.%
\footnote{ A straightforward way of seeing this is by replacing the transverse
Gaussian by a plane wave, $\varphi_{0}(\rho)=\frac{1}{\sqrt{2\pi a_{\perp}}}$,
$\rho<a_{\perp}/2$: Then the 1D interaction would have the form $D^{2}/\sqrt{a_{\perp}^{2}+\rho^{2}}^{3}$. %
}

\begin{figure}
\begin{centering}
\includegraphics[width=0.9\columnwidth]{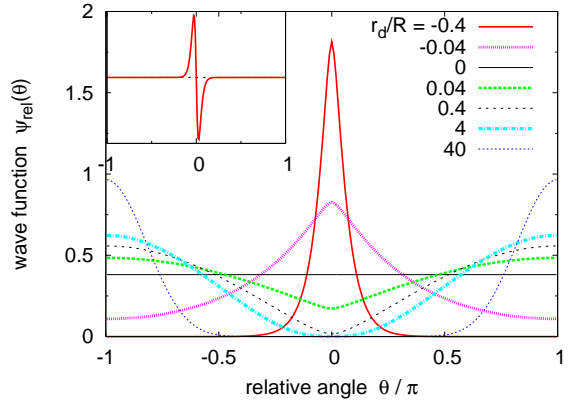}
\par\end{centering}

\caption{\emph{(}color online) Relative wave function $\psi(\vartheta)$ of
two bosonic dipoles on a ring. For weak repulsive interactions, $r_{d}/R<1$,
the $\psi$ resembles that of bosons with contact interactions. For
stronger coupling, $r_{d}/R>1$, the wave function becomes localized
at $\vartheta=\pm\pi$. For sufficient attraction, the bosons form
a bound state. \emph{Inset}: For identical \emph{fermions}, a tightly
bound state  may form for stronger attraction (here $r_{d}/R=-4$).
\label{fig:rel-wavefn}}
\end{figure}

As an illustration, we have diagonalized $h_{\mathrm{rel}}$ numerically
and plotted the ground-state wave function $\psi_{\mathrm{rel}}(\vartheta)$
in Fig.~\ref{fig:rel-wavefn} for different $r_{d}/R$. For \emph{bosons},
the noninteracting wave function is the zero-momentum state $\psi(\vartheta)=\frac{1}{\sqrt{2\pi}}$.
For $r_{d}/R\ll1$, the dominant contribution comes from the peak
of $V_{\mathrm{rel}}$ at $\vartheta=0$, where the probability amplitude
is reduced. With increasing $r_{d}$, the minimum value of $\psi(0)$
becomes deeper until it reaches down to almost zero (see, e.g., $r_{d}/R=0.4$),
in which case $\psi(\vartheta)\sim|\sin\frac{\vartheta}{2}|$ resembles
the modulus of the wave function of noninteracting \emph{fermions}.
This behavior is reminiscent of bosons with a short-range (contact)
interaction.

For $r_{d}/R\gtrsim1$, the weaker $|r|^{-3}$ tail becomes effective,
driving the two dipoles apart until their density is sharply peaked
at a distance $|\vartheta|=\pi$ (i.e, the dipoles are at opposite
poles of the ring, $|r|=2R$). This is a precursor of a crystal-like
state. For fermions, the modulus $|\psi(\vartheta)|$ would look similar. 

In the spirit of our comment above, let us consider an effective attraction,
$r_{d}<0$. Figure~\ref{fig:rel-wavefn} illustrates that two bosons
form a bound state, which here becomes localized for $|r_{d}|/R\gtrsim0.01$.
This is analogous to the well-known bound state of a 1D delta interaction.
However, even identical fermions can form a {}``p-wave'' bound state
(Fig.~\ref{fig:rel-wavefn}, inset), provided the potential is deep
enough to accommodate a second, anti-symmetric level. Note that it
owes its existence to the non-zero range $\sim a_{\perp}$ of the
potential. 

Let us mention that the relative state for $N=2$ fermions (or any
even number) on a ring is a subtle issue, as can be seen from the
Slater determinant $(e^{\pm i\theta_{2}}-e^{\pm i\theta_{1}})\propto e^{\pm i\frac{\theta_{1}+\theta_{2}}{2}}\sin\frac{\theta_{1}-\theta_{2}}{2}$.
The CM state has nonzero (angular) momentum $\pm1$, whereas the relative
orbital, having momentum $\frac{1}{2}$, is \emph{anti}periodic in
$\vartheta\equiv\theta_{1}-\theta_{2}$ in order to preserve $2\pi$-periodicity
of the total wave function. This relates to the fact that any even-$N$
Fermi sea is twofold degenerate, and thus will either have nonzero
momentum or, in the case of a symmetric Fermi-level state $e^{ik_{F}\theta_{N}}+e^{-ik_{F}\theta_{N}}\propto\cos(k_{F}\theta_{N})$,
a cosine modulation of the CM wave function, breaking the translational
symmetry.

\section{Homogeneous case ($\alpha=0$) \label{sec:Homogeneous}}

Before tackling inhomogeneous interactions, let us first focus on
the homogeneous case, $\alpha=0$. Here the dipoles are oriented perpendicular
to the plane of the ring; thus, rotational symmetry is preserved and
the interaction is independent of $\Theta$, $V_{\mathrm{1D}}(\theta_{1},\theta_{2})=D^{2}V_{\mathrm{rel}}(\theta_{1}-\theta_{2})/4a_{\perp}^{3}$.
This system therefore has a close analogy to a linear 1D system, with
particle distances $x_{i}-x_{j}\equiv2R\sin(\frac{\theta_{1}-\theta_{2}}{2})$.
In order to exhibit the essential physics, we first discuss ground-state
properties for a few particles as obtained using the numerically exact
multi-configurational time-dependent Hartree method (see Appendix
\ref{sec:method}). We then proceed to discuss analytical models for
the limiting cases of gas-like and solid-like regimes.

\subsection{Ground-state properties}

\begin{figure}
\begin{centering}
\includegraphics[width=0.6\columnwidth]{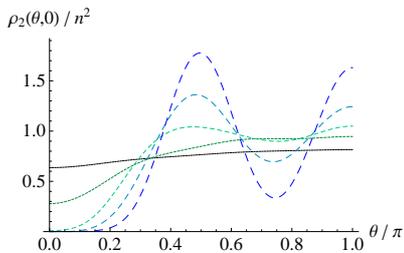}
\par\end{centering}

\caption{(color online) Pair-distribution function $\rho_{2}(\theta,0)$ in
the homogeneous case $\alpha=0$ ($4$ bosons), for $nr_{d}=0.0025$,
$0.025$, $0.25$, $2.5$ and $25.5$ (with increasing dashing)\emph{.
\label{fig:rho2_alpha0}}}
\end{figure}

Since the number density $n(\theta)=\langle\sum_{i}\delta(\theta-\theta_{i})\rangle=N/2\pi$
is constant for any $nr_{d}$ due to translational invariance in $\theta$,
we focus on two-particle correlations. These are described, e.g.,
by the pair distribution function $\rho_{2}(\theta,\theta')=\sum_{i\neq j}\langle\delta(\theta-\theta_{i})\delta(\theta'-\theta_{j})\rangle$,
which gives the probability density of finding one dipole at $\theta$
and a second one at $\theta'$. This is plotted in Fig.~\ref{fig:rho2_alpha0}
for $N=4$ \emph{bosons}. 

For dipole lengths much smaller than the inter-particle distance,
$nr_{d}\ll1$, the initially uniform pair distribution develops a
{}``correlation hole'' at $\theta=\theta'$ with increasing coupling
--- indicating that finding two dipoles at the same position becomes
more and more unlikely --- until the pair distribution strongly resembles
that of noninteracting fermions (see, e.g., $nr_{d}=0.25$) exhibiting
Friedel oscillations. This gas-like behavior is reminiscent of bosons
interacting through a short-range or contact interaction $g\delta(x-x')$
\cite{lieb63a}, an observation similarly made in the context of a
dipolar gas in a harmonic trap \cite{deuretzbacher09}. Throughout
this regime, \emph{fermions }would remain essentially noninteracting
due to the Pauli principle. 

As the dipolar length becomes comparable with the mean inter-particle
separation, $nr_{d}\gtrsim1$, the long-range tail of the interaction
becomes crucial. Then the pair distribution develops more pronounced
oscillations, which culminate in well-separated peaks for $nr_{d}\gg1$.
These indicate the crystal-like localization of the individual particles
\cite{arkhipov05,deuretzbacher09}.

\begin{figure}
\begin{centering}
\includegraphics[width=0.8\columnwidth]{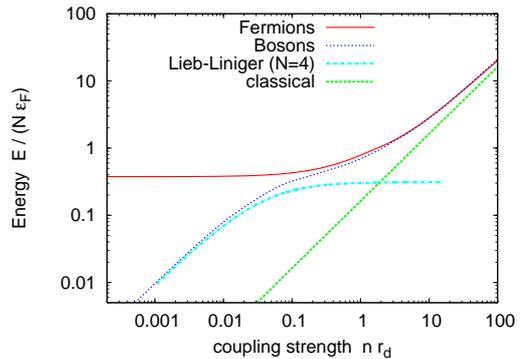}
\par\end{centering}

\caption{(color online) Ground-state energy $E(nr_{d})$ for $N=4$ dipoles
in the homogeneous system ($\alpha=0$). Also plotted for comparison:
the energy of $N$ Lieb-Liniger bosons (with $g=D^{2}/a_{\perp}^{2}$)
and the classical limit $E_{C}$.\emph{ }\label{fig:energy-alpha0}}
\end{figure}

\subsection{Bosons: Lieb-Liniger-gas regime ($nr_{d}\ll1$)}

For weak dipolar coupling, $r_{d}\ll1/n$, we observe a behavior reminiscent
of a 1D Bose gas with short-range interactions. Let us now model this
gas-like regime.

The 1D Bose gas is described by the Lieb-Liniger model \cite{lieb63a}
\begin{equation}
H_{LL}=-\frac{\hbar^{2}}{2m}\sum_{i=1}^{N}\frac{\partial^{2}}{\partial x_{i}^{2}}+g\sum_{i<j}\delta(x_{i}-x_{j}),\label{eq:LiebLiniger}\end{equation}
where here $x_{i}\equiv R\theta_{i}$. In the thermodynamic limit
$N,R\to\infty$ (keeping the density $n=N/2\pi R$ fixed), the properties
of the Lieb-Liniger model are completely determined by the dimensionless
parameter \cite{lieb63a} \[
\gamma=\frac{m}{\hbar^{2}n}g\equiv-\frac{2}{na_{1}}>0,\]
where $a_{1}=-2\hbar^{2}/mg$ denotes the 1D scattering length. More
specifically, the Lieb-Liniger gas exhibits a crossover from a weakly
interacting Bose gas for $\gamma\ll1$ to a so-called Tonks gas in
the {}``fermionization'' limit $\gamma\gg1$, similar to our findings
in the gas-like regime. We will now delineate a relation between the
model parameter $\gamma$ (i.e., $g$) and the 1D dipolar interaction
(parametrized by $r_{d},a_{\perp}$).

\subsubsection{Born approximation for the interaction}

In the spirit of the Born approximation, one may identify the coupling
constant with the zero-momentum Fourier transform of the potential,
$g\approx V_{1D}(q=0)=\lim_{q\to0}\int dx\, V_{1D}(x)e^{iqx}$. For
$V_{1D}(x)\equiv D^{2}V_{\mathrm{rel}}(x/a_{\perp})/4a_{\perp}^{3}$,
this yields \begin{equation}
g=\frac{D^{2}}{2a_{\perp}^{2}}\int_{0}^{\infty}\negthickspace ds\, V_{\mathrm{rel}}(s)=\frac{D^{2}}{a_{\perp}^{2}}\quad\Longrightarrow\quad\gamma=\frac{r_{d}}{2na_{\perp}^{2}}.\label{eq:Born}\end{equation}

To check this, in Fig.~\ref{fig:energy-alpha0} we have plotted the
ground-state energy $E(nr_{d})$ obtained numerically for dipolar
interactions as well as for the Lieb-Liniger model. To compare the
two on an equal footing, we have made use of the results for $N=4$
bosons based on the solution of the $N$-particle Lieb-Liniger equations
\cite{sakmann05} (rather than the integral equations valid in the
thermodynamic limit), with $g$ given by (\ref{eq:Born}). For $\gamma\ll1$,
i.e., $nr_{d}\ll2(na_{\perp})^{2}\sim0.01$, the system is well described
by a weakly interacting Bose gas, $E\simeq\frac{N-1}{2}ng$. Conversely,
for $2(na_{\perp})^{2}\ll nr_{d}\ll1$, one has $\gamma\gg1$, and
the bosons repel each other so strongly that they become virtually
impenetrable. For very tight confinement, $na_{\perp}\to0$, the bosons
are thus practically fermionized already at very low coupling, which
is the case for a purely $1/|r|^{3}$ potential \cite{arkhipov05}. 

In the strict limit $g\to+\infty$ ($a_{1}\to0^{-}$), the Lieb-Liniger
state $\Psi_{a_{1}}$ is closely connected to that of non-interacting
identical fermions $\Psi_{F}$ via the so-called Bose-Fermi map \cite{girardeau60},
\[
\Psi_{a_{1}\to0^{-}}=\mathcal{A}\Psi_{\mathrm{F}};\quad\mathcal{A}(x_{1}\dots x_{N})\equiv\prod_{i<j}\mathrm{sgn}(x_{i}-x_{j}).\]
Note that for the ground state, $\Psi_{0}=|\Psi_{\mathrm{F}}|$. Consequently,
only phases differ, whereas all local quantities are identical. Thus
the pair distribution function (Fig.~\ref{fig:rho2_alpha0}) displays
a correlation hole of size $r\sim1/k_{F}$, $k_{F}=\pi n$ being the
Fermi wave number, and the energy approaches $E\to\sum_{|k|\le k_{F}}\frac{(\hbar k)^{2}}{2m}\stackrel{N\gg1}{\simeq}\frac{1}{3}N\epsilon_{F}$,
where $\epsilon_{F}=(\hbar k_{F})^{2}/2m$. Notice, however, that
here the fermionization limit is slightly subtle. First, $\gamma$
in (\ref{eq:Born}) cannot tend to infinity so long as $nr_{d}\ll1$;
rather, at $\gamma\text{\ensuremath{\gtrsim}}1/2(na_{\perp})^{2}$,
nonzero-range effects come into play which are no longer described
by the Lieb-Liniger model. Moreover, for even numbers $N=2,4,\dots$,
the Bose-Fermi map is not applicable on a ring \cite{girardeau60},
since the Fermi gas has CM momentum $k_{F}=\pi n$ for reasons discussed
in Sec.~\ref{sub:Relative-problem}. Thus, the Tonks gas has an energy
lower than that of the Fermi gas by the CM energy $(\hbar k_{F})^{2}/2Nm$.
Generally, for any finite $N$, the energy is slightly lower than
the thermodynamic-limit result $E=\frac{1}{3}N\epsilon_{F}$ by a
factor of $(N^{2}-1)/N^{2}$.

\subsubsection{Beyond the Born approximation}

Let us now discuss the validity of the Born approximation (\ref{eq:Born})
and how to include effects beyond it.

In one dimension, the Born approximation holds for 1D scattering lengths
large compared with the scattering wavelength, $|ka_{1}|\gg1$, and
becomes exact in the non-interacting limit $a_{1}\to-\infty$. This
can be derived straightforwardly from the Born expansion of the T
matrix. For low-energy scattering, $ka_{\perp}\ll1$, this leads to
the condition $|a_{1}|\gg a_{\perp}$. Since for the dipolar interaction
we have $|a_{1}|\simeq a_{\perp}^{2}/r_{d}$,\emph{ }this is equivalent
to \emph{ \[
r_{d}\ll a_{\perp}.\]
 }Thus, the Lieb-Liniger map (\ref{eq:Born}) is valid for $nr_{d}\ll na_{\perp}$,
tightening the constraint found above. As an illustration, for $N=4$
and $R/a_{\perp}=10$, we have $na_{\perp}\approx0.064$, i.e., (\ref{eq:Born})
holds up to $\gamma\sim1/(na_{\perp})$.

So, what corrections for $g$ do we have to expect for stronger couplings,
$r_{d}\gtrsim a_{\perp}$? Although that regime is hard to realize
experimentally, the question is of theoretical interest. To answer
it, we have calculated numerically the 1D scattering length, $a_{1}=-2\hbar^{2}/mg$,
for the dipolar potential $V_{1D}$. This is done by solving the $E=0$
Schrödinger equation for the relative motion, $\left[-\frac{\hbar^{2}}{m}\frac{d^{2}}{dr^{2}}+V_{1D}(r)\right]\psi(r)=0$,
and fitting it to the asymptotic low-energy form $\psi(r)\simeq c(|r|-a_{1})$
for $|r|\gg a_{\perp}$. 

\begin{figure}
\begin{centering}
\includegraphics[width=0.8\columnwidth]{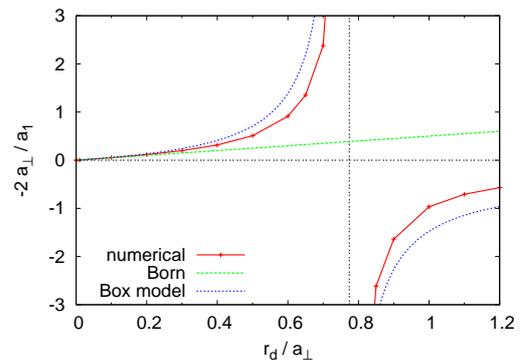}
\par\end{centering}

\caption{(color online) Effective 1D interaction strength $g\equiv-2\hbar^{2}/ma_{1}$
(units of $\hbar^{2}/ma_{\perp}$) for the dipolar potential $V_{1D}$
as a function of $r_{d}/a_{\perp}$. For comparison: Born approximation
$g\approx D^{2}/a_{\perp}^{2}$, and results for a box potential with
corresponding parameters. \label{fig:a1_Born}}
\end{figure}

The results are shown in Fig.~\ref{fig:a1_Born}. Clearly, the Born
approximation remains qualitatively correct up until $r_{d}/a_{\perp}\approx0.5$,
corresponding to $nr_{d}\approx0.5na_{\perp}\approx0.032$ above.
However, as $r_{d}/a_{\perp}\to0.77\dots$, a resonance appears, signifying
that $g\to\infty$, and then $g$ crosses over to $g\to-\infty$.
This resonance explains why, in Fig.~\ref{fig:energy-alpha0}, the
fermionization limit was reached already much earlier than naively
expected from (\ref{eq:Born}). Moreover, it suggests an interpretation
for energies larger than that of the Tonks gas, but still in the gaseous
regime: A negative value of $g$ indicates the existence of a so-called
Super-Tonks state \cite{astrakharchik05}, which exhibits correlations
analogous to a hard-core potential with nonzero range $a_{1}>0$.\textbf{
}Before discussing this in more depth, let us illustrate briefly why
the resonance comes about.

\paragraph{Box-potential model}

To obtain some intuition for interpreting the numerical results obtained
above, let us consider a crude model for the dipolar interaction above:
a box potential $V(x)=v\Theta(\ell-|x|)$ of range $\ell\sim a_{\perp}$
and height $v\sim D^{2}/a_{\perp}^{3}$. The 1D scattering length
is known analytically, \[
\frac{a_{1}}{\ell}=1-\frac{\coth\eta}{\eta},\quad\eta\equiv\sqrt{\frac{mv}{\hbar^{2}}}\ell\sim\sqrt{\frac{r_{d}}{a_{\perp}}}.\]
Of course, for $\eta\ll1$, the Born approximation is recovered, $a_{1}/\ell\simeq-1/\eta^{2}$,
or $g\simeq2\ell v\sim D^{2}/a_{\perp}^{2}$. However, at $\eta=\eta^{\star}\approx1.2$
(i.e., for $r_{d}$ of order $a_{\perp}$), $a_{1}$ vanishes like
$a_{1}/\ell\simeq\frac{1}{2}(\eta^{2}-\eta^{\star2})$, so that $g\propto(r_{d}-r_{d}^{\star})^{-1}$
exhibits a resonance at the critical point. This is exactly the type
of behavior encountered in Fig.~\ref{fig:a1_Born}.

To understand this phenomenon a little more deeply, it is worth considering
what happens to the wave function. For $\eta\ll1$, the (negative)
scattering length is much larger than the box radius, $|a_{1}|\gg\ell$,
which reflects that the length scale of the wave function is too large
to sample any details of the potential. In other words, the typical
energy is so small compared with that associated with the potential's
structure, $h^{2}/ma_{1}^{2}\ll\hbar^{2}/m\ell^{2}$, that the physics
is basically shape independent and entirely determined by the Born
parameter. However, when $\eta\sim1$, we have $|a_{1}|\lesssim\ell$
, i.e., the essential change occurs within the interior of the box,
where the wave function is suppressed more and more strongly. In the
extreme case of a hard-sphere potential, $\eta\gg1$, $\psi(r)=0$
for all $|r|<\ell$, so that the scattering length $a_{1}=\ell>0$
(in the dipolar case, $a_{1}\to+\infty$). By continuity, $a_{1}$
must change sign in between.

\paragraph{Super-Tonks regime}

The results above permit some analytical statements about the regime
where $a_{1}>0$, i.e., beyond the resonance. There the system can
be understood as a gas of 1D hard spheres (rods) with diameter $\ell=a_{1}$.
As this resembles a so-called Tonks gas of fermionized bosons, with
diameter $a_{1}\to0^{-}$, this is sometimes referred to as the Super-Tonks
regime \cite{astrakharchik05,astrakharchik08}. 

In a homogeneous system, the wave function for $na_{1}>0$ can be
obtained from the fermionized wave function ($a_{1}=0$) via the mapping\textbf{
}\cite{nagamiya40} \[
\Psi_{a_{1}}\left(\{x_{j}\}\right)=\Psi_{0^{-}}\left(\{x_{j}-(j-1)a_{1}\}\right),\]
 where ordering $x_{j}<x_{j+1}-a_{1}$ is understood, and the total
length $L$ entering $\Psi_{0}$ is replaced by $L-Na_{1}$. With
that substitution, the hard-sphere energy can be obtained from the
fermionized value, $E_{a_{1}\to0}=N(\hbar\pi n)^{2}/6m$, \begin{equation}
E_{a_{1}}=\frac{E_{0}}{(1-na_{1})^{2}}.\label{eq:STG}\end{equation}

In principle, one might expect this formula to also give a good description
of the dipolar ground-state energy $E(nr_{d})$ when inserting $a_{1}(r_{d})$
as obtained from the scattering calculation, at least in the neighborhood
of $na_{1}=0$. However, in our case this regime is not well separated
from the long-range one, $nr_{d}\gg1$, as explained above. Consequently,
we do not find Eq.~(\ref{eq:STG}) to yield a reasonable agreement
for our parameters.

\subsection{Localization ($nr_{d}\gg1$)}

Let us now discuss the limiting case $nr_{\mathrm{d}}\gg1$. We found
that the long-range dipolar repulsion becomes dominant here, so that
the particles localize. Another indicator of this is the ground-state
energy of the dipoles (Fig.~\ref{fig:energy-alpha0}), which deviates
from the saturated fermionization regime and changes over to an energy
increase proportional to $r_{\mathrm{d}}$ for $nr_{d}\gg1$. In this
regime, the dipoles become classical in the sense that the potential
energy overwhelms the kinetic energy, so that for $nr_{d}\to\infty$
the ground state becomes a position eigenstate $|\bar{\theta}_{1},\dots,\bar{\theta}_{N}\rangle_{\pm}$
($\pm$ indicating bosonic/fermionic permutation symmetry), where
the angles $\{\bar{\theta}_{j}\}$ are determined by minimizing the
energy \[
E_{C}=\sum_{i<j}\frac{D^{2}}{\left|2R\sin\frac{\bar{\theta}_{i}-\bar{\theta}_{j}}{2}\right|^{3}}.\]
This yields a crystal-like, equidistant distribution of particles:
$\bar{\theta}_{j}=j\frac{2\pi}{N}+\delta$, $j\in\{0,\dots,N-1\}$,
which is only defined up to an angle $\delta$ by rotational invariance,
and an energy $E_{C}=\frac{ND^{2}}{16R^{3}}\sum_{\nu=1}^{N-1}\left|\sin\frac{\nu\pi}{N}\right|^{-3}$.
To find the behavior in the thermodynamic limit, one may approximate
the sum by an integral, $\sum_{\nu}\left|\sin\frac{\nu\pi}{N}\right|^{-3}\approx2\frac{N}{\pi}\int_{\pi/2N}^{\pi/2}d\phi\left|\sin\phi\right|^{-3}\simeq4N^{3}/\pi^{3}$
as $N\to\infty$. In that limit, we thus have $E_{C}/N\approx2nr_{d}\times(\hbar n)^{2}/2m$,
an expression formally similar to that of an infinite line \cite{arkhipov05}. 

However, for finite $nr_{d}$, the particles are not strictly localized.
Rather, their wave packets are spread out over a width $w$ due to
their zero-point motion, as is illustrated by the pair distribution
function in Fig.~\ref{fig:rho2_alpha0}. We now estimate $w$. For
simplicity, we consider a linear 1D system and perform a classical
normal-mode analysis: Expanding the interaction potential $D^{2}\sum_{i<j}|x_{i}-x_{j}|^{-3}$
to second order about the minimum (a lattice with nearest-neighbor
distance $1/n$) yields effective harmonic-oscillator frequencies
$\omega_{C}=\sqrt{12D^{2}n^{5}/m}$, corresponding to an oscillator
width $w\equiv\sqrt{\hbar/m\omega_{C}}$ given by \[
nw=\sqrt[4]{\frac{1}{12nmD^{2}/\hbar^{2}}}\sim\frac{1}{\sqrt[4]{nr_{\mathrm{d}}}}.\]
 This confirms that for dipole lengths large compared to the average
inter-particle spacing, $nr_{\mathrm{d}}\gg1$, the dipoles should
indeed be well localized, in agreement with Fig.~\ref{fig:rho2_alpha0}.
Taking into account the zero-point motion about the classical equilibrium
positions, we find that the crystal energy $E_{C}$ is corrected by
$+N\hbar\omega_{C}/2\sim N(\hbar n)^{2}/m\times\sqrt{nr_{d}/2}$.

In the crystalline limit $nr_{d}\to\infty$ where the behavior is
nearly classical, the same reasoning applies to fermions. For finite
coupling, however, due to the Pauli principle, the kinetic energy
of fermions is inherently higher than that of bosons. In this sense,
stronger repulsion is needed to suppress the zero-point motion compared
to bosons, so that the crystalline limit is approached more slowly.

\section{Repulsive interactions ($0<\alpha<\alpha_{\mathrm{c}}$)\label{sec:Repulsion}}

Let us now consider the case where the tilt angle is nonzero such
that the effective interaction is still repulsive but inhomogeneous.
We first present the ground-state properties obtained for the few-particle
system, before discussing simple analytic models for the gaseous and
crystal-like limits.

\subsection{Ground-state properties}

\begin{figure}
\begin{centering}
\includegraphics[width=0.8\columnwidth]{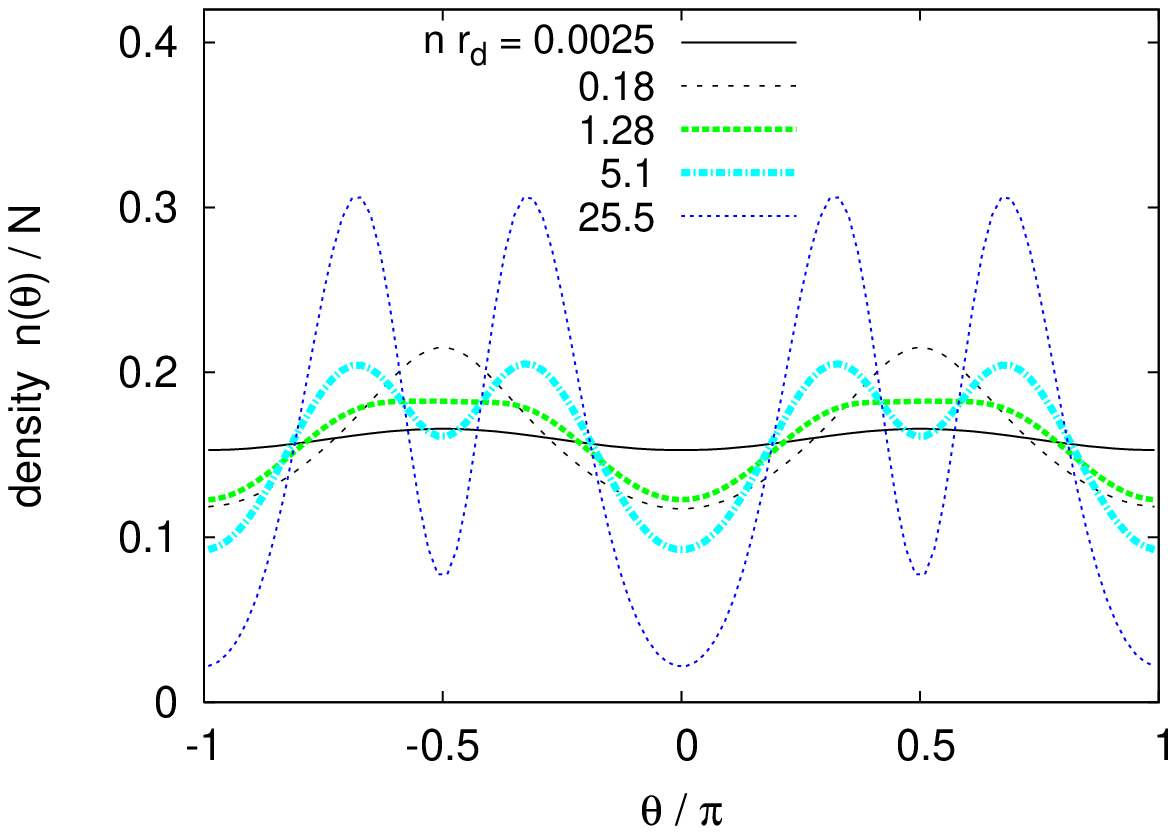}
\par\end{centering}

\begin{centering}
\includegraphics[height=0.3\columnwidth]{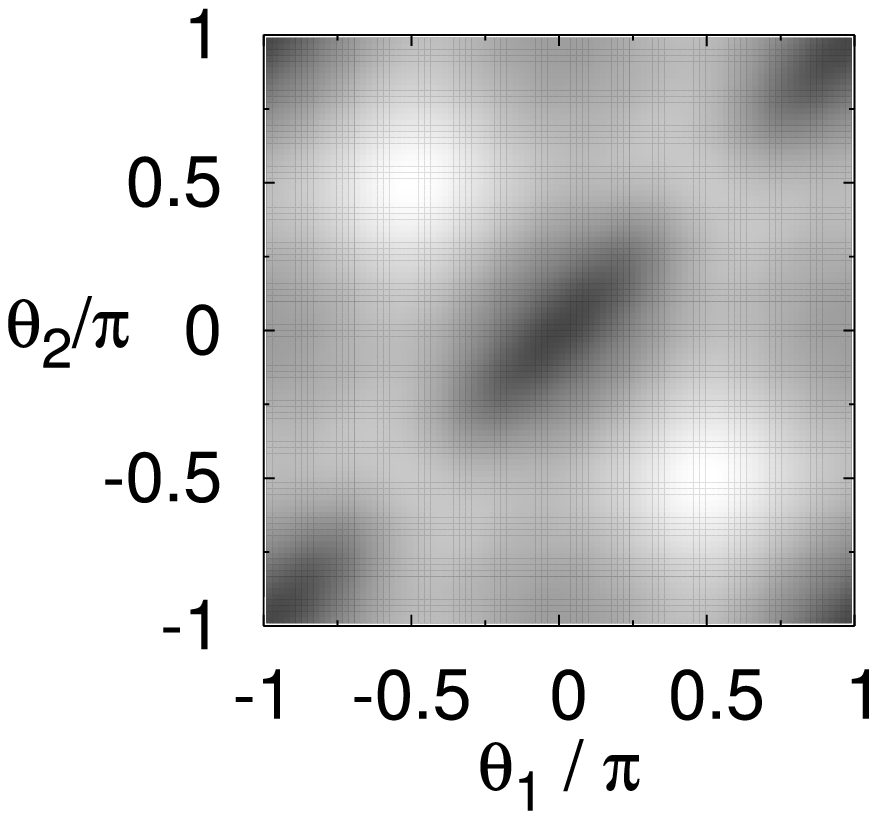}\includegraphics[height=0.3\columnwidth]{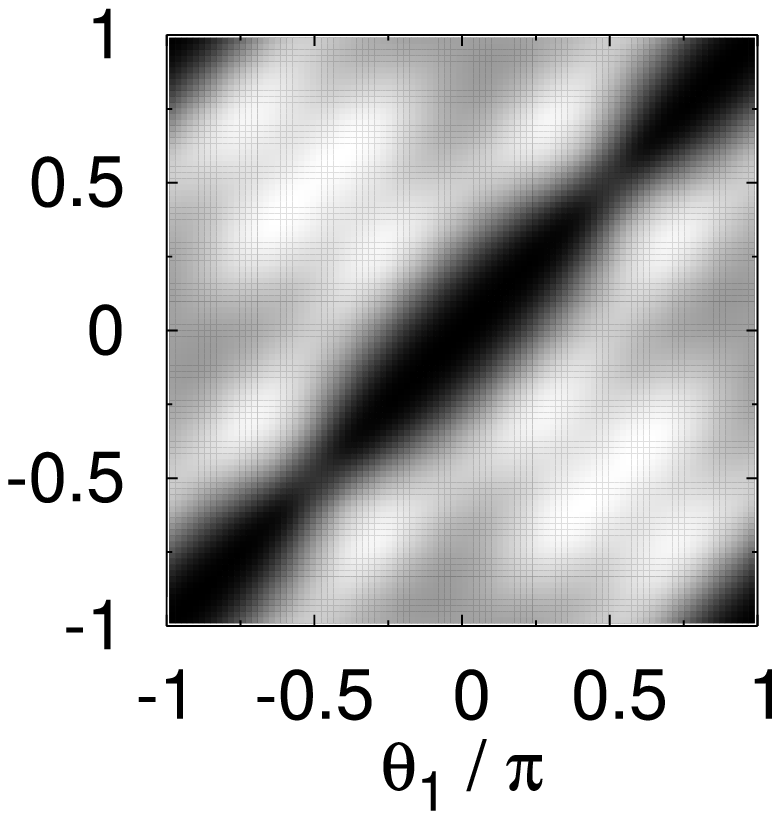}\includegraphics[height=0.3\columnwidth]{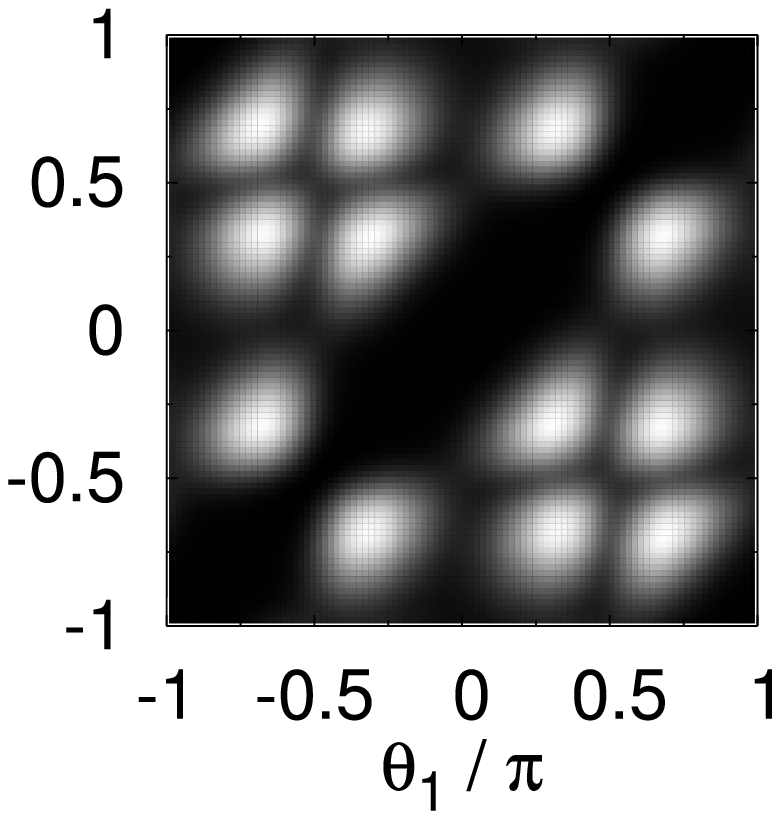}
\par\end{centering}

\caption{(color online) Inhomogeneous case ($\alpha=0.19\pi$\emph{, }$N=4$
bosons): Density profile $n(\theta)$ {[}top{]} and pair distribution
function $\rho_{2}(\theta_{1},\theta_{2})$ {[}below{]} for $nr_{d}=0.025,1.28,25.5$
(left to right).\textbf{\emph{ \label{fig:density_alpha0.19}}}}
\end{figure}

To bring out the essential effect, we first focus on bosons with an
inhomogeneity just below the threshold value, $\alpha=0.19\pi\lesssim\alpha_{\mathrm{c}}$.
In contrast to dipoles aligned perpendicular to the plane of the ring,
for $\alpha\neq0$ the system is no longer rotationally invariant
about $\theta$. Thus interaction effects are reflected in the number
density $n(\theta)$, as shown for $N=4$ bosons in Fig.\ \ref{fig:density_alpha0.19}.
For small couplings, $nr_{d}\ll1$, the state resembles that of a
Bose gas. The  density has a slightly stronger weight near the potential
minima $\theta=\pm\pi/2$ for intermediate couplings but tends to
flatten out for larger values, $nr_{d}\sim1$. By contrast, for $nr_{d}\gg1$,
the density profile develops $N=4$ distinct peaks, revealing a crystal-like
localization of the particles. Note that the peak positions are not
equidistant, but rather slightly displaced toward $\theta=\pm\pi/2$
due to the pronounced minima of $V_{\mathrm{CM}}$.%
\footnote{For an \emph{odd }number of dipoles, localization is more subtle.
For, say, $N=3$, there are two degenerate ground states classically
-- one would thus see a symmetry-averaged $2N=6$ peaks in the density
$n(\theta)$, rather just $N=3$.%
} 

The transition from a gas-like to a localized state is also clearly
visible in the pair distribution function $\rho_{2}(\theta,\theta')$
shown in Fig.~\ref{fig:density_alpha0.19}: For $nr_{d}\ll1$, two-body
correlations are absent in the mean-field regime, $\rho_{2}(\theta_{1},\theta_{2})\approx\rho(\theta_{1})\rho(\theta_{2})$,
or are limited to Friedel-like oscillations around the correlation
hole near $\theta_{1}=\theta_{2}$, reminiscent of a fermionized Bose
gas. For $nr_{d}\gg1$, the pinning of the particles to individual
peaks is clearly discernible.

A similar trend would be observed for fermions: For $nr_{d}\ll1$,
we find a gas-like state, which however is only weakly interacting
owing to the Pauli exclusion principle. With increasing coupling,
$nr_{d}\gtrsim1$, the state crosses over into a crystal-like one
characterized by localization of the particles in the classical minimum-energy
configuration, just as in the Bose case. Due to the Fermi energy,
that crossover is much smoother than for the bosons.

\subsection{Inhomogeneous Lieb-Liniger regime ($nr_{d}\ll1$)}

Let us dwell on the parameter range where the ground state is gas-like.
In the homogeneous limit $\alpha\to0$, we established that here the
system is well described by the Lieb-Liniger Hamiltonian (\ref{eq:LiebLiniger}),
with the 1D interaction strength $g=D^{2}/a_{\perp}^{2}$ (for $r_{d}\ll a_{\perp}$).
By extension, for $\alpha\neq0$, the system should resemble a Bose
gas with an \emph{inhomogeneous} interaction $g\left(\frac{x_{1}+x_{2}}{2}\right)\delta(x_{1}-x_{2})$,
where $g(x)=D^{2}(1-3\sin^{2}\alpha\sin^{2}\frac{x}{R})/a_{\perp}^{2}$
is the Fourier transform of $V_{1D}$ at zero relative momentum. 

Although this inhomogeneous Lieb-Liniger Hamiltonian system is not
integrable, some insight can be gained by assuming slow variation
of $g(x)$ compared with the local correlation length. In this case,
a local-density approximation \cite{oliva88} may be applied, which
consists in assuming the equation of state of the homogeneous Lieb-Liniger
system, $\mu=f_{g}(n)$, to hold \emph{locally,} $\mu=f_{g(x)}[n(x)]$.
Here $\mu$ denotes the chemical potential as a function of number
density $n(x)$; moreover, for the Lieb-Liniger system, $f_{g}(n)$
can be obtained directly from the ground-state energy, which is known
in terms of the dimensionless function $e(\gamma)=E/N\frac{(\hbar n)^{2}}{2m}$
\cite{lieb63a,dunjko01}. Note that the chemical potential is $x$
independent, as local equilibrium is assumed to hold. Carrying out
the local-density approximation numerically for the full range of
$\gamma$ is beyond the present scope. However, considering the borderline
case also provides some insight into the impact of the inhomogeneity.

\subsubsection{Mean-field regime, $\gamma(x)\ll1$}

For locally \emph{weak }interactions, $\gamma\ll1$, the homogeneous
equation of state reads $\mu=ng$, which yields a profile \[
n(x)=\frac{\mu}{g(x)}=\frac{n(0)}{1-3\sin^{2}\alpha\sin^{2}\frac{x}{R}}.\]
 Remarkably, this is independent of $\gamma$, and thus of $nr_{d}$,
since $\mu\mbox{\ensuremath{\propto}}g$, that is, kinetic-energy
contributions are negligible in the homogeneous equation of state.
This holds only for large enough particle numbers $N\gg1$ and not
too large variations of $g(x)$; otherwise the kinetic pressure will
smear out the density gradients on the scale of the coherence length
or smaller. 

It is worth mentioning that the local-density approximation for $\gamma\ll1$
corresponds to the Thomas-Fermi approximation, which neglects the
kinetic energy in the Gross-Pitaevskii equation for the mean-field
orbital $\phi(x)$ \cite[Ch. 6]{pethick}, \[
-\frac{\hbar^{2}}{2m}\phi''(x)=\left[\mu-n(x)g(x)\right]\phi(x)\approx0.\]

While the strict mean-field limit assumes all bosons to be condensed
into a single orbital, $\Psi=\phi^{\otimes N}/N^{N/2}$, even for
weak repulsion a deviation from a fully uncorrelated state occurs.
Conveniently, this can be quantified, e.g., in the second-order correlation
function, $g_{2}(x,x)\equiv\rho_{2}(x,x)/n(x)^{2}$. In the homogeneous
case, $g_{2}=1-\frac{2}{\pi}\sqrt{\gamma}+O(\gamma)$ \cite{gangardt03}.
Replacing $\gamma=mg/n\hbar^{2}$ by $\gamma(x)=mg(x,x)/n(x)\hbar^{2}\simeq mg^{2}(x,x)/\hbar^{2}\mu$
yields $g_{2}(x,x)\simeq1-\frac{2}{\pi}\sqrt{\gamma(0)}(1-3\sin^{2}\alpha\sin^{2}\frac{x}{R})$,
amounting to a pronounced suppression of the pair distribution $\rho_{2}(\theta,\theta)$
near $\theta\equiv x/R=0$, as observed in Fig.~\ref{fig:density_alpha0.19}.

\subsubsection{Tonks-gas regime, $\gamma(x)\gg1$}

For stronger coupling, the density modulation (Fig.~\ref{fig:density_alpha0.19},
$nr_{d}=1.3$) is attenuated until, in the fermionization limit $\gamma(x)\gg1$,
$n(x)$ approaches a constant. This is clear since, by the Pauli principle,
identical fermions do not sense any contact interaction and hence
no spatial modulation of $g(x)$. Formally, this can be seen from
the chemical potential in the Tonks regime, $\mu=(\hbar\pi n)^{2}/2m$,
which becomes independent of $g$. Since $\mu$ is spatially constant
by assumption, so is $n(x)=n(0)$. 

For large but finite $g$, pair correlations are actually not fully
suppressed, $g_{2}\simeq0+\frac{4}{3}(\pi/\gamma)^{2}$. Since $\gamma(x)\simeq mg(x)/n\hbar^{2}$,
this implies fairly strong residual pair correlations near the potential
minima $\theta=\pm\pi/2$, $g_{2}(x)\simeq\frac{4\pi^{2}}{3\gamma(0)^{2}}\left(1-3\sin^{2}\alpha\sin^{2}\frac{x}{R}\right)^{-2}$.
Moreover, as discussed for the homogeneous case, the limit $\gamma(x)\gg1$
corresponds to fermionization only if $nr_{d}\ll1$, locally. Instead,
for increasing $nr_{d}$, we find that the incipient fermionization
gives way to a state where the nonzero range of the repulsive potential
becomes relevant, until, eventually, a crystal-like localization occurs.

\subsection{Inhomogeneous crystal-like regime ($nr_{d}\gg1$)}

We now present analytical models for the crystal-like limit $nr_{d}\gg1$.
Similarly to the case $\alpha=0$ , here each dipole is pinned to
a position $\bar{\theta}_{i}$ determined by the extrema of the classical
energy \[
E_{C}=\frac{D^{2}}{8R^{3}}\sum_{i<j}\frac{1-3\sin^{2}\alpha\sin^{2}\frac{\bar{\theta}_{i}+\bar{\theta}_{j}}{2}}{\sin^{3}\left|\frac{\bar{\theta}_{i}-\bar{\theta}_{j}}{2}\right|}.\]
 In this case, though, the equilibrium configuration is not simply
an equidistant lattice. Rather, owing to the competition of potential
energy of the relative motion (favoring maximum distance) and the
CM (alignment near $\theta=\pm\pi/2$), the dipoles are localized
at discrete positions which are distributed \emph{inhomogeneously}
on the ring. In principle, the distribution of $\{\bar{\theta}_{i}\}$
can be found by numerical minimization of $E_{C}$. However, to obtain
some analytic insight, we will do this for one of the simplest nontrivial
cases, $N=4$. Complementing this few-body viewpoint, we then extend
the local-density approximation above to the strong-coupling regime.

\emph{}

\subsubsection{Few-particle model}

Let us consider the special case $N=4$. By symmetry, we can write
$\bar{\theta}_{i}=\pm(\frac{\pi}{2}\pm\phi)$, $i=1,\dots,4$. The
one-dimensional potential-energy curve \[
\frac{E_{C}(\phi)}{D^{2}/4R^{3}}=\frac{1-3\sin^{2}\alpha}{\sin^{3}\phi}+\frac{1}{\cos^{3}\phi}+\left(1-3\sin^{2}\alpha\sin^{2}\phi\right)\]
consists of three contributions:
\begin{enumerate}
\item The first term represents the weak repulsion of a pair at, say, $\frac{\pi}{2}\pm\phi$,
which has minimum CM energy ($\Theta=\pi/2$). 
\item The second term represents the energy of a configuration of type $\pm(\frac{\pi}{2}+\phi)$,
which has maximum repulsion ($\Theta=0$). 
\item The remaining term is that of two diametrically opposed dipoles, e.g.,
at $\pm\frac{\pi}{2}+\phi$; here \emph{only }the slowly varying CM
potential contributes.
\end{enumerate}
The total potential consists mainly of (1.) a steep potential barrier
at $\phi\to0$ (repulsion of a pair near $\pm\pi/2$) and (2.) a confinement
for larger $\phi$ due to inter-pair repulsion, adding up to a potential
well for $\phi$. Neglecting the third term, the potential minimum
can be found to be at \begin{equation}
\phi_{*}(\alpha)=\arctan\sqrt[5]{1-3\sin^{2}\alpha}.\label{eq:minimum-angle}\end{equation}
 Near $\alpha=0$, this reproduces the uniform lattice spacing $2\pi/N$,
with a very smooth correction for $\alpha>0$, $2\phi_{*}(\alpha)\simeq\frac{\pi}{2}-\frac{3}{5}\alpha^{2}$.
It is only very close to the critical angle that a noticeable inhomogeneity
appears, $\phi_{*}(\alpha)\simeq\sqrt[5]{1-3\sin^{2}\alpha}\simeq8^{5/8}\sqrt[5]{\alpha_{c}-\alpha}$.
At $\alpha\to\alpha_{c}^{-}$, this goes to zero very abruptly, indicating
an attempted clustering as the minimum energy vanishes. Including
the slightly deconfining third term above, the minimum separation
angle is marginally shifted toward higher $\phi$. Moreover, although
we chose $N=4$ as a special soluble case, the basic competition between
bunching toward $\theta=\pm\pi/2$ and keeping a minimum inter-particle
distance will persist for the case of more than a single pair, $N>4$. 

Let us comment on the limit $\alpha\to\alpha_{c}$. The non-analyticity
of $\phi_{*}(\alpha)$ at $\alpha_{c}$ hints at the fact that, at
least for $nr_{d}\to\infty$, there is a phase transition when $\alpha$
crosses the critical angle. In fact, we will see in Sec.~\ref{sec:Attraction}
that for $\alpha>\alpha_{c}$ bound states appear which are tightly
localized at $\theta=\pm\pi/2$ for $nr_{d}\to\infty$. For $\alpha<\alpha_{c}$,
however, the two dipoles at, say, $\pi/2$ cannot get arbitrarily
close due to the quantum-mechanical kinetic energy neglected so far,
contrary to what we would expect based on (\ref{eq:minimum-angle}).

\subsubsection{Local-density approximation}

Complementary to the few-particle model discussed above, let us consider
the limit $N\to\infty$. If the CM potential varies on a length scale
much larger than the average inter-particle distance, we may assume
that, locally, the system behaves similarly to a homogeneous 1D system.
Then the local-density approximation may be applied, i.e., that the
equation of state of the homogeneous system, $\mu=f_{r_{d}}(n)$,
holds locally, $\mu=f_{r_{d}(\Theta)}[\bar{n}(\Theta)]$. Here $\bar{n}(\Theta)$
is the average number density around $\Theta$, and $r_{d}(\Theta)=r_{d}\times(1-3\sin^{2}\alpha\sin^{2}\Theta)$.

If we take the chemical potential of the \emph{linear }homogeneous\emph{
}system for $nr_{d}\to\infty$, $\mu=4\zeta(3)\hbar^{2}n^{3}r_{d}/m$
\cite{arkhipov05}, we find \[
\bar{n}(\Theta)\propto\frac{1}{\sqrt[3]{1-3\sin^{2}\alpha\sin^{2}\Theta}}.\]
 This density variation is rather smooth due to the strong dipolar
repulsion. For smaller $\alpha$, the density is practically homogeneous,
with tiny humps near $\Theta=\pm\pi/2$. For $\alpha\to\alpha_{c}^{-}$,
though, the density becomes markedly peaked and tends toward $\delta(\Theta\mp\frac{\pi}{2})$,
indicating an instability toward clustering beyond $\alpha_{c}$.
Of course, the local-density approximation is valid only for density
variations slow compared with the local coherence length: Close to
the critical angle, $\bar{n}(\Theta)$ is smeared out on the length
scale given by the zero-point motion.

\section{Partially attractive interactions ($\alpha_{\mathrm{c}}<\alpha<\alpha_{c2}$)
\label{sec:Attraction}}

For tilt angles $\alpha>\alpha_{c}$, the interaction potential is
no longer purely repulsive but acquires attractive regions for CM
angles $|\Theta|=\frac{\pi}{2}\pm\frac{\Delta\Theta}{2}$ (Fig.~\ref{fig:potential}),
where \[
\Delta\Theta=2\arccos\left(\frac{\sin\alpha_{c}}{\sin\alpha}\right)\equiv\frac{\Delta x}{R}.\]
Note that for $\alpha\to\alpha_{c}^{+}$, this attractive segment
is vanishingly narrow, $\Delta\Theta\simeq2^{7/4}\sqrt{\alpha-\alpha_{c}}$,
whereas for $\alpha\to\frac{\pi}{2}$, it makes up almost two thirds
of the total circumference, $\Delta\Theta/\pi\to0.608\dots$ . With
increasing coupling, the competition between repulsion and attraction
leads to an intriguing crossover from gas-like states toward clustered
{}``droplets'', which we will first discuss on the basis of few-body
simulations, before giving an analytic description of the cluster
states.

\subsection{Crossover from gas-like to clustered states}

\begin{figure}
\begin{centering}
\includegraphics[width=0.8\columnwidth]{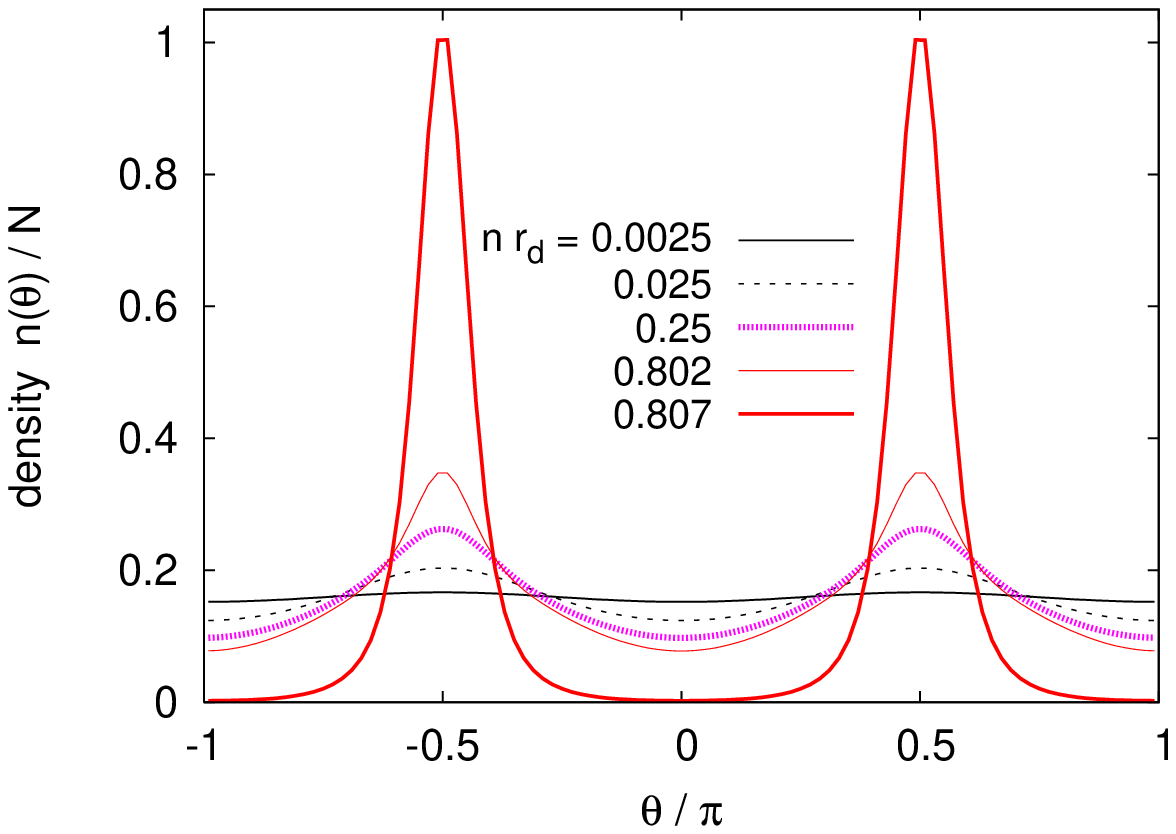}
\par\end{centering}

\begin{centering}
\includegraphics[height=0.3\columnwidth]{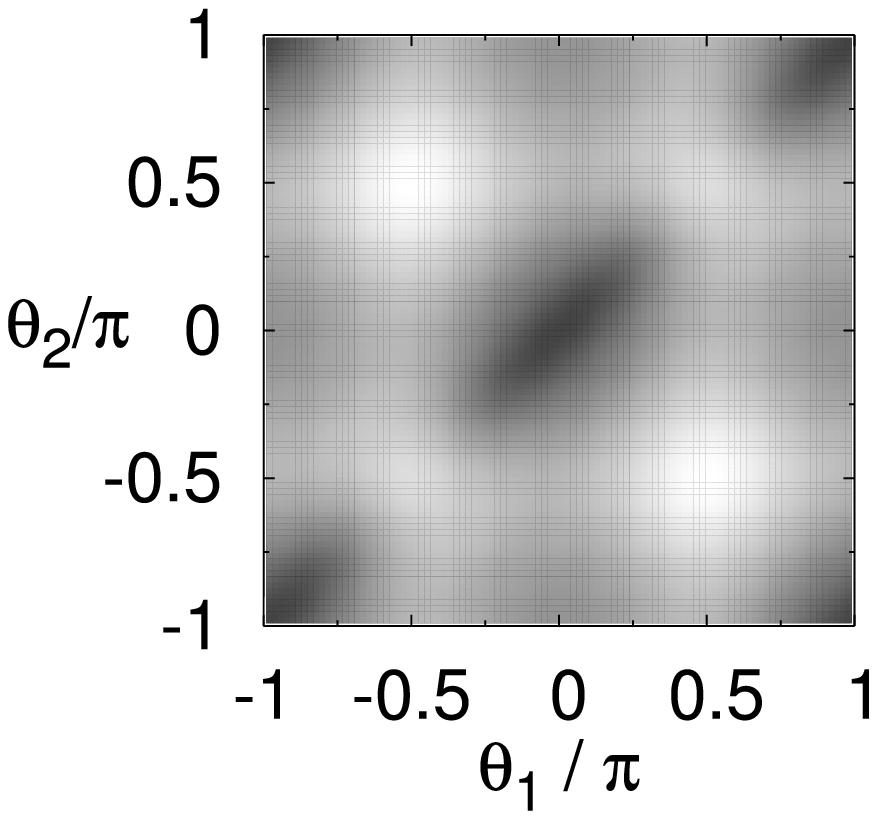}\includegraphics[height=0.3\columnwidth]{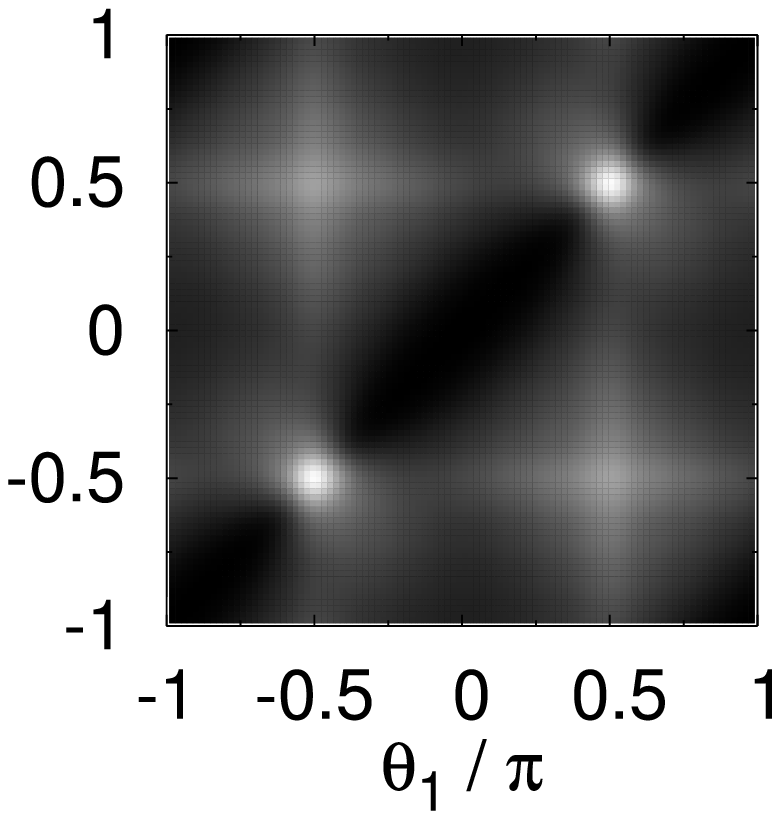}\includegraphics[height=0.3\columnwidth]{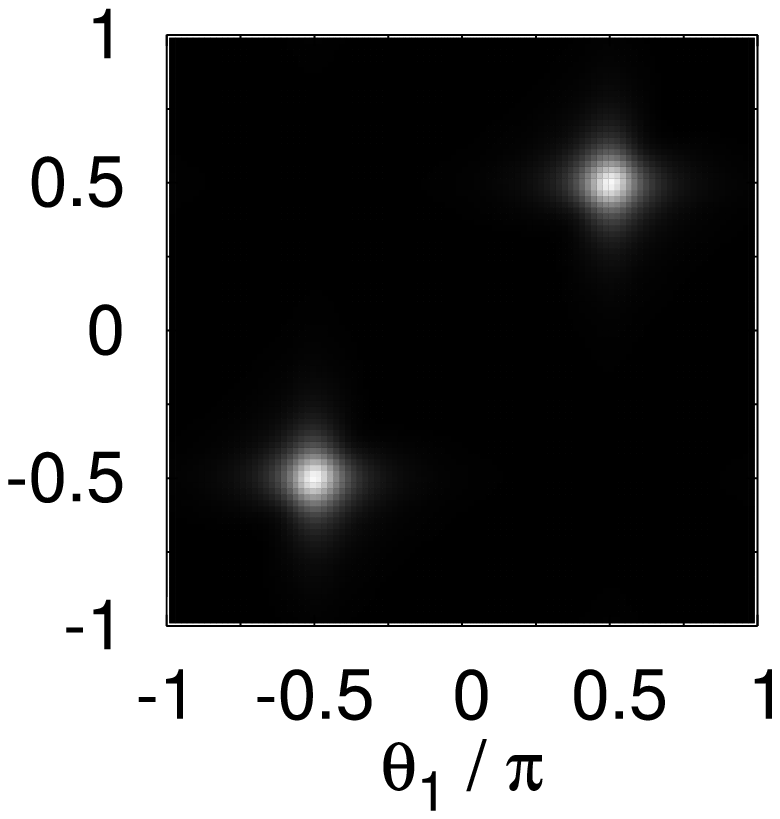}
\par\end{centering}

\caption{(color online) Clustering of $N=4$ bosons ($\alpha=0.2\pi$): Density
profile $n(\theta)$ {[}top{]} and pair distribution function $\rho_{2}(\theta_{1},\theta_{2})$
{[}below{]} for $nr_{d}=0.025,\,0.802$ and $0.807$ (left to right).
\label{fig:density_alpha0.2}}
\end{figure}

\begin{figure}
\begin{centering}
\includegraphics[width=0.8\columnwidth]{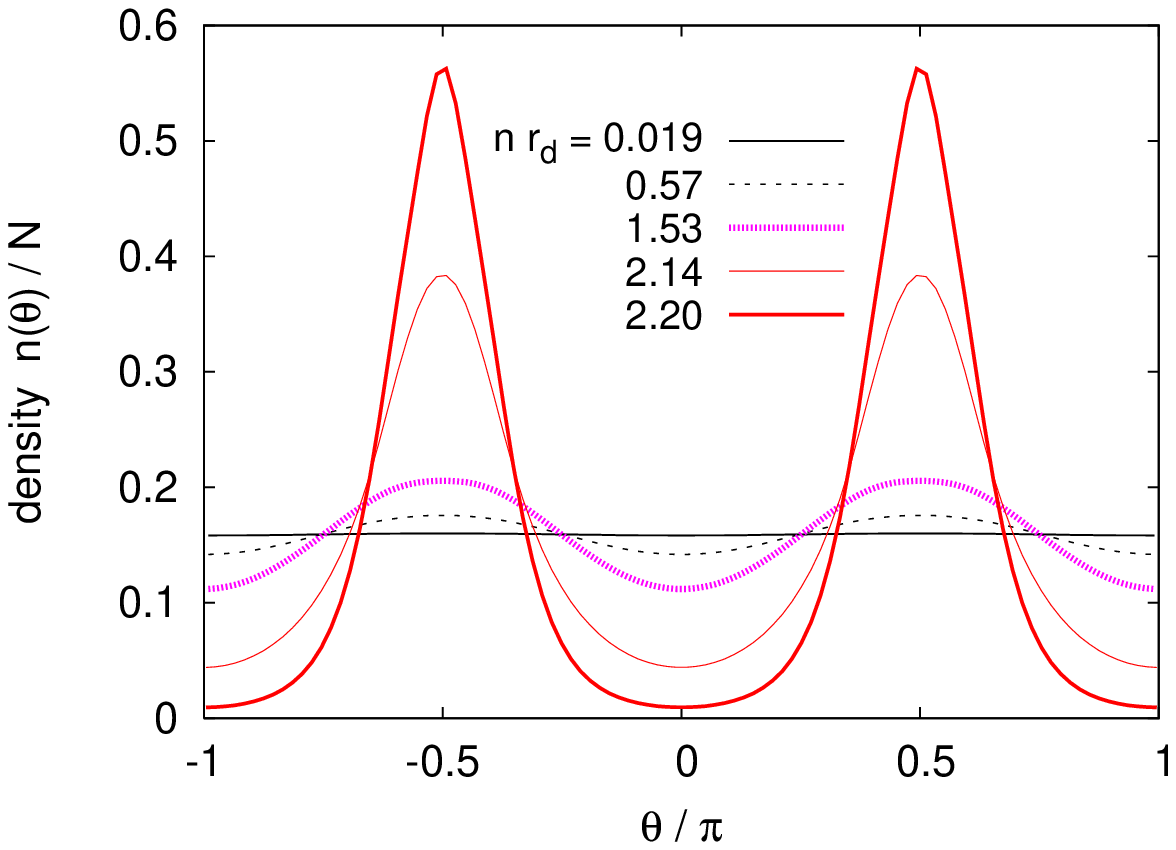}
\par\end{centering}

\begin{centering}
\includegraphics[height=0.3\columnwidth]{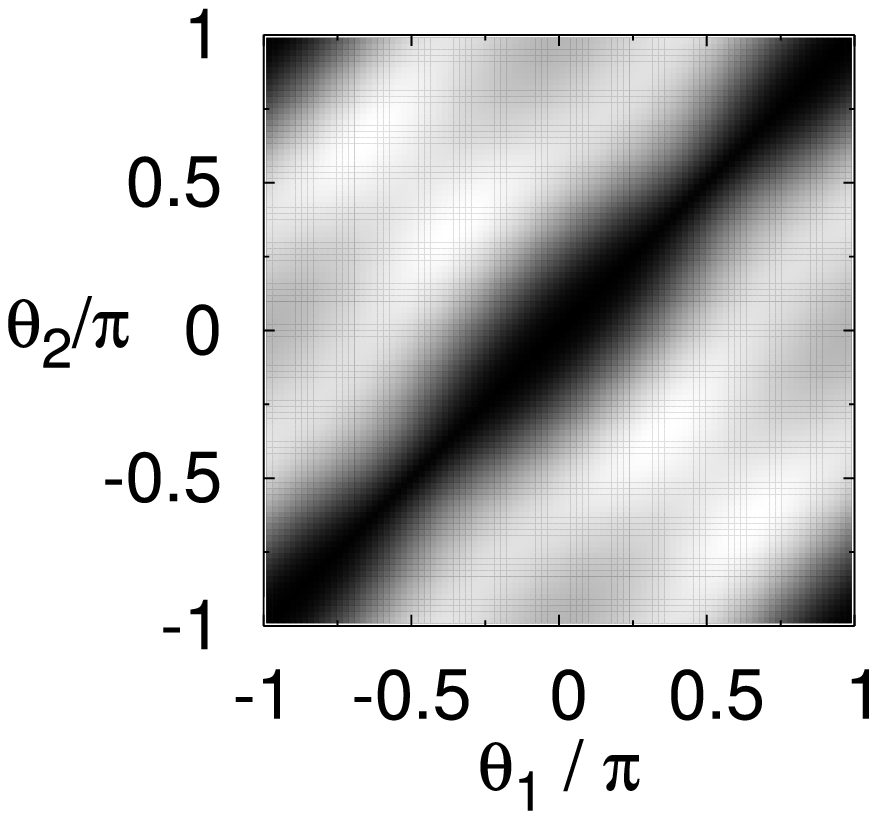}\includegraphics[height=0.3\columnwidth]{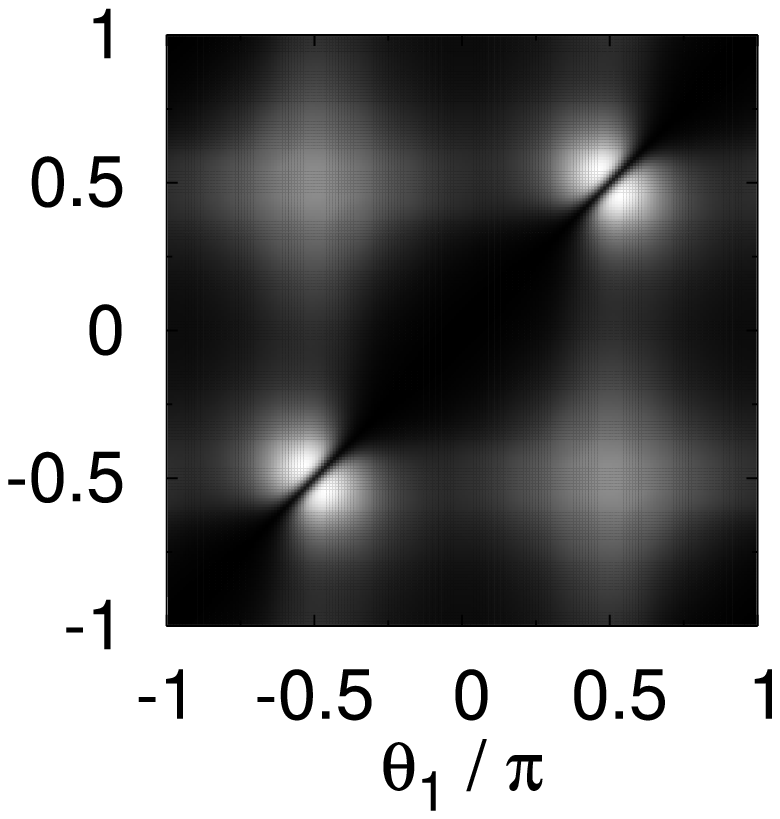}\includegraphics[height=0.3\columnwidth]{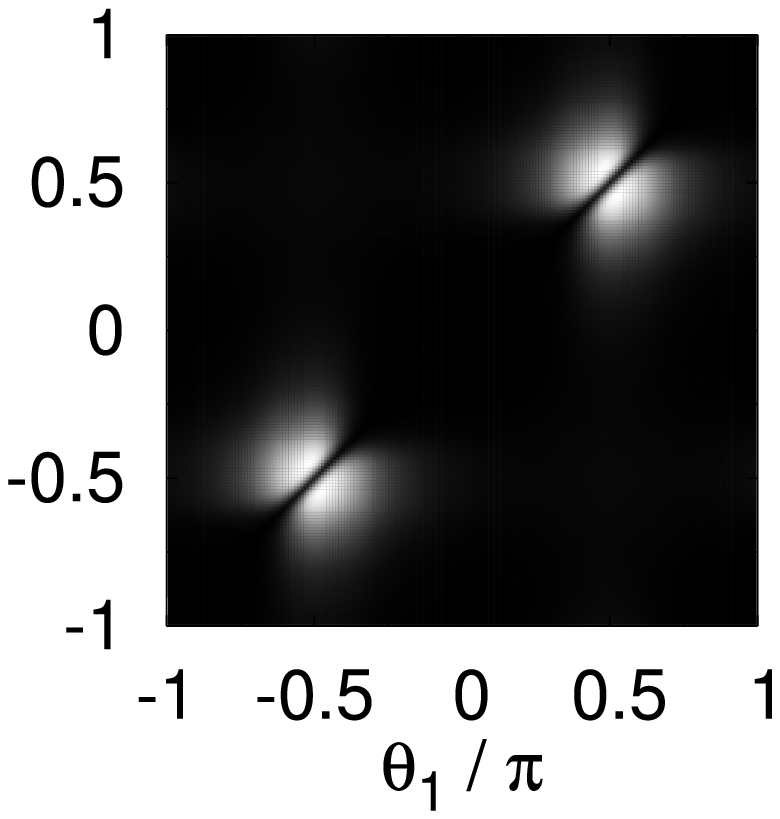}
\par\end{centering}

\caption{(color online) Clustering of $N=3$ fermions ($\alpha=0.22\pi$):
Density profile $n(\theta)$ {[}top{]} and pair distribution function
$\rho_{2}(\theta_{1},\theta_{2})$ {[}below{]} for $nr_{d}=0.57,\,2.14$
and $2.20$ (left to right) .\textbf{\label{fig:fermi-density_alpha0.22}}}
\end{figure}

Let us first illustrate the crossover from weak to strong coupling
by means of the density $n(\theta)$, shown in Fig.~\ref{fig:density_alpha0.2}
($N=4$ bosons, $\alpha=0.2\pi$) and Fig.~\ref{fig:fermi-density_alpha0.22}
($N=3$\textbf{ }fermions, $\alpha=0.22\pi$). In both cases, the
densities go over from a homogeneous profile, characteristic of a
gas-like state, to one with two sharp peaks, indicating cluster formation. 

\begin{figure}
\begin{centering}
\includegraphics[width=0.8\columnwidth]{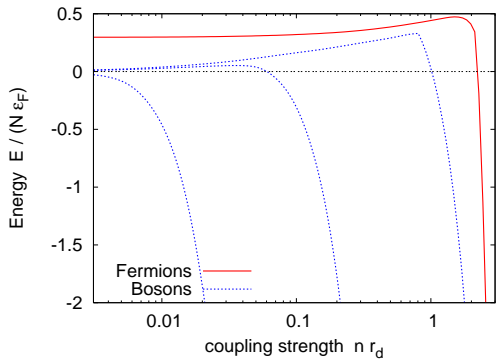}
\par\end{centering}

\caption{(color online) Ground-state energy $E(nr_{d})$ in the case of partial
attraction: $N=3$ fermions (---, $\alpha=0.22\pi$); $N=4$ bosons
(- - -, $\alpha=0.2\pi,\,0.22\pi$, and $\pi/2$ from top to bottom).\emph{
\label{fig:energy-alpha0.22}}}
\end{figure}

It is noteworthy that in these examples, the transition to the localized
clustered state is very sharp (e.g., at $nr_{d}\approx0.8$ for the
bosons). This is also reflected in the ground-state energy $E(nr_{d})$,
shown in Fig.~\ref{fig:energy-alpha0.22}. In fact, whereas the energy
eventually falls and becomes negative at high enough $nr_{d}$ for
any $\alpha>\alpha_{c}$, the behavior for small coupling crucially
depends on $\alpha$: Closer to the critical angle, the bosonic energy
increases for small coupling, and it is only beyond a critical value
of $nr_{d}$ that it starts decreasing rather abruptly. By contrast,
for larger $\alpha$, $E(nr_{d})$ is a monotonically decreasing function.
To understand that competition between repulsion and attraction, note
that---for $nr_{d}\to0$---the (bosonic) energy behaves as \begin{eqnarray}
E & \simeq & \frac{N(N-1)}{2}\langle V_{\mathrm{CM}}(\Theta)\rangle_{0}\left\langle V_{\mathrm{rel}}(\vartheta)\right\rangle _{0}\label{eq:MFshift}\\
 & \sim & \frac{N}{2}ng(0)\left(1-{\scriptstyle \frac{3}{2}}\sin^{2}\alpha\right),\nonumber \end{eqnarray}
where the average is over the (uniform) noninteracting CM and relative
states. Since $g(0)>0$, the energy increase for small $nr_{d}$ is
\emph{positive} for \[
\alpha<\alpha_{c2}\equiv\arcsin\sqrt{\frac{2}{3}}\approx0.304\pi.\]
 In other words, for tilt angles less than $\alpha_{c}$, the potential
is predominantly repulsive. Initially, the delocalized bosons will
thus undergo an analogous evolution as in Sec.~\ref{sec:Repulsion},
e.g., behave like an inhomogeneous Bose gas, which is predominantly
repulsive ($\alpha<\alpha_{c2}$) or has a net attraction ($\alpha>\alpha_{c2}$).
For identical fermions, the interaction-energy shift per particle
will be negligible compared with the Fermi energy for all $\alpha$.

This makes clear that the transition to a clustered state also strongly
depends on $\alpha$. Beyond $\alpha_{c2}$, there is a smooth crossover
from an attractive gaseous state to a clustered one (Fig.~\ref{fig:energy-alpha0.22})
at rather small coupling. By contrast, for only slightly overcrtical
$\alpha\in(\alpha_{c},\alpha_{c2})$ -- e.g., $\alpha=0.2\pi$ in
Fig.~\ref{fig:density_alpha0.2} -- the transition takes place at
much larger coupling and is more subtle. With increasing $nr_{d}$,
the repulsive gas-like state first goes over into a transitional state
with strong fragmentation (cf. $nr_{d}=0.802$), which can be thought
of roughly as a Mott-insulator-type state $|\frac{N}{2},\frac{N}{2}\rangle$
with $N/2$ dipoles at each of the poles, $\theta=\pm\pi/2$. This
can be best seen from the pair distribution $\rho_{2}(\theta,\theta')$,
Fig.~\ref{fig:density_alpha0.2}(center), which upon closer inspection
confirms that a measurement of one particle at, say, $\pi/2$ yields
a measurement of $(\frac{N}{2}-1)$ particles at $\pi/2$ and $\frac{N}{2}$
at $-\pi/2$. There is a sharp avoided crossing to a clustered state
($nr_{d}=0.807$), with all $N$ particles localized at either $\pm\pi/2$.
By parity symmetry, the exact ground state is actually a superposition
or Schrödinger-cat state, $\frac{1}{\sqrt{2}}\left(|N,0\rangle+|0,N\rangle\right)$,
as is evident from the pair distribution function in Fig.~\ref{fig:density_alpha0.2}.
A small symmetry-breaking perturbation will likely cause a collapse
of the superposition.

For fermions, the critical interaction strength needed for cluster
formation is higher (Fig.~\ref{fig:energy-alpha0.22}). This is because
of the Pauli principle: The exchange hole discernible in $\rho_{2}$
(Fig.~\ref{fig:fermi-density_alpha0.22}) strongly reduces the average
attraction between two fermions. Moreover, due to the Pauli pressure,
the fermions are more spread out spatially and require a stronger
attraction to be squeezed into the narrow zone of attraction. 

To obtain a deeper understanding of cluster formation, including its
dependence on the particle number and on $\alpha$, we will now develop
a simplified analytical model that can reproduce its basic features.

\subsection{Simple cluster model: Homogeneous case\label{sub:Cluster-hom}}

As a first step, let us consider the case of a homogeneous system
with purely attractive interaction. This is interesting in its own
right, since clustered states are not limited to a ring geometry.
Moreover, as we will see below, this also allows for a qualitative
understanding of the inhomogeneous case.

\subsubsection{Bosons}

Consider $N$ bosonic dipoles interacting with a homogeneous attractive
contact interaction $\bar{g}<0$. (On the ring, we may identify this
with the value of $g(x)$ in the minimum $x_{0}/R=\pm\pi/2$, $\bar{g}=D^{2}(1-3\sin^{2}\alpha)/a_{\perp}^{2}<0$.)
Let us assume all bosons to be localized at some $x_{0}$, spread
out over a length $L$, e.g., by making a variational ansatz with
all particles in a single orbital, $\phi_{L}(x)=e^{-(x-x_{0})^{2}/2L^{2}}/\sqrt[4]{\pi L^{2}}$.
Then, provided $L\gg a_{\perp}$, the interaction energy scales as
$E_{{\rm int}}\sim N^{2}\bar{g}/L$, whereas $E_{{\rm kin}}\sim N\hbar^{2}/mL^{2}$.
For $\bar{g}<0$, the total energy $E(L)$ has a local minimum, which
yields the size of an equilibrium cluster, \begin{equation}
L\sim\frac{\bar{a}_{1}}{N},\quad\bar{a}_{1}\equiv-\frac{2\hbar^{2}}{m\bar{g}}.\label{eq:binding-length}\end{equation}
 The characteristic length in the two-body case is, of course, the
local 1D scattering length $\bar{a}_{1}>0$, and larger $N$ lead
to a shrinking of the cluster. The cluster's binding energy is then
\[
E\sim-N^{3}\frac{\hbar^{2}}{m\bar{a}_{1}^{2}}\sim-N^{3}\frac{\hbar^{2}|r_{d}(\frac{\pi}{2})|^{2}}{ma_{\perp}^{4}},\]
 which diverges quadratically with the coupling strength. 

Note that in the language of the Lieb-Liniger model, this can be understood
as the well-known {}``bright-soliton'' ground state of attractive
1D bosons \cite{mcguire64,castin01a}: \[
E=-\frac{N(N^{2}-1)}{6}\frac{\hbar^{2}}{m\bar{a}_{1}^{2}};\quad\Psi(\boldsymbol{x})\propto e^{-\sum_{i<j}|x_{i}-x_{j}|/\bar{a}_{1}}.\]
 Since $E/N\propto N^{2}$, the binding energy is unbounded in the
thermodynamic limit $N\equiv\bar{n}L\to\infty$, and the state is
stable only for a small enough number of particles. 

In free space, this cluster size would be extremely large for weak
attraction. On the ring, however, for the cluster to form, it must
fit into the attractive regions of length $\Delta x=2R\arccos(1/\sqrt{3}\sin\alpha)$
around $\theta=\pm\pi/2$, \begin{equation}
L<\Delta x.\label{eq:cluster-criterion_Ldx}\end{equation}
 This yields a rough estimate for the critical interaction strength
necessary for cluster formation, \begin{equation}
nr_{d}\gtrsim\frac{\left(a_{\perp}/R\right)^{2}}{\pi\arccos\left(\frac{1}{\sqrt{3}\sin\alpha}\right)|1-3\sin^{2}\alpha|},\label{eq:cluster-criterion_nrd}\end{equation}
 which qualitatively reproduces the parameters necessary for cluster
formation found numerically \cite{zoellner10a}. \textbf{}

\subsubsection{Fermions }

We now discuss the fermionic case. As above, we assume the fermions
to be in an essentially noninteracting state, but with a localization
length $L$. Then the kinetic energy, $E_{{\rm kin}}=N(\hbar k_{F})^{2}/6m=N^{3}\pi^{2}\hbar^{2}/6mL^{2}$,
with $k_{F}\equiv\pi N/L$, has the same structure as in the bosonic
case except for a factor of order $N^{2}$ due to the exclusion principle.
By contrast, the interaction energy has a more subtle form since the
nonzero range of the potential is crucial. Setting again $D^{2}(1-3\sin^{2}\alpha)/a_{\perp}^{2}\equiv\bar{g}<0$
and $\rho_{2}(x,x')=\left(\frac{N}{L}\right)^{2}g_{2}(x-x')$, we
can write \begin{equation}
E_{\mathrm{int}}=\frac{N^{2}\bar{g}}{L}\int_{0}^{L}\frac{dr}{2a_{\perp}}\, g_{2}(r)V_{\mathrm{rel}}\left(\frac{r}{a_{\perp}}\right).\label{eq:Eint-Fermi}\end{equation}
 For an ideal Fermi gas, \[
g_{2}(r)=1-\left[\frac{\sin(\frac{\pi N}{L}r)}{N\sin(\frac{\pi}{L}r)}\right]^{2}\stackrel{N\gg1}{\simeq}1-\left[\frac{\sin(k_{F}r)}{k_{F}r}\right]^{2}.\]
 There is no general analytic formula for the integral (\ref{eq:Eint-Fermi}),
but we can make statements about limiting cases. First, as $L\ll a_{\perp}$
tends to zero, the integral gives $LV_{\mathrm{rel}}(0)/2a_{\perp}$,
so \[
E_{\mathrm{int}}\to\frac{N^{2}\bar{g}}{2a_{\perp}}\qquad(L\to0).\]
Being bounded, the interaction energy is thus completely outweighed
by the kinetic energy ($\propto L^{-2}$), which prevents the collapse
toward high-density clusters. On the other hand, for large sizes,
$L\gg Na_{\perp}$, the dominant contribution comes from the intermediate
zone $a_{\perp}\ll r\ll L/\pi N$ where $V_{\mathrm{rel}}(s)\simeq4/s^{3}$
and $g_{2}(r)\simeq\frac{1}{3}(k_{F}r)^{2}$, and \[
E_{\mathrm{int}}\sim N^{3}\bar{g}a_{\perp}^{2}\frac{\ln(L/\pi Na_{\perp})}{L^{3}}\qquad(L\to\infty).\]
This falls off slightly faster with $L$ than the kinetic energy.
Hence, fermions cannot form a bound state with arbitrarily large size.
This reflects the fact that, for low densities, the attractive interaction
energy will be largely canceled by the exchange term. However, binding
is possible for an intermediate regime, $a_{\perp}\ll L\ll Na_{\perp}$.
Here the density is high enough so that the direct (Hartree) interaction
term is large, $g_{2}(r)\approx1$, but not so high for binding to
be suppressed by the kinetic pressure:\[
E_{\mathrm{int}}\approx\frac{N^{2}}{2L}\bar{g}\qquad(a_{\perp}\ll L\ll Na_{\perp}).\]

It is only for such moderate sizes $L$ that clusters may exist. Under
this assumption, the minimum energy is found straightforwardly for
\begin{equation}
L\approx\frac{\pi^{2}}{3}N\bar{a}_{1}.\label{eq:binding-length-Fermi}\end{equation}
 By the Hartree approximation, this has the same structure as for
the Bose cluster, but here the cluster becomes \emph{larger} for higher
particle number $N$ because of the Pauli principle. Note that this
is valid only under two constraints: 

(i) $L\ll Na_{\perp}$ for the Hartree term to be dominant, i.e.,
$\bar{a}_{1}\ll a_{\perp}$ or $|r_{d}(\frac{\pi}{2})|\gg a_{\perp}$.
Minimizing $E(L)$ numerically, we find the onset of clustering at
$|r_{d}(\frac{\pi}{2})|\gtrsim a_{\perp}$, which relaxes the condition.

(ii) $|\frac{\partial E}{\partial N}|\ll\hbar^{2}/ma_{\perp}^{2}$
for the strictly 1D description to hold. This is equivalent to $\bar{a}_{1}\gg a_{\perp}$,
the complementary condition to the Hartree limit (i). However, since
the latter criterion can be relaxed considerably, it is plausible
that Fermi clustering may be found in a regime where the physics is
no longer strictly 1D. This may require further study, going beyond
the bare single-mode 1D description.

The discussion so far assumed a 1D system with a homogeneous attraction.
On the ring, the cluster must furthermore fit into the attractive
regions, $L<\Delta x$ (\ref{eq:cluster-criterion_Ldx}). This yields
the estimate \begin{equation}
nr_{d}\gtrsim\frac{\frac{2}{3}\pi^{3}\left(na_{\perp}\right)^{2}}{\arccos\left(\frac{1}{\sqrt{3}\sin\alpha}\right)|1-3\sin^{2}\alpha|},\label{eq:cluster-criterion_nrd-Fermi}\end{equation}
 which essentially differs from the bosonic one (\ref{eq:cluster-criterion_nrd})
by a factor of order $N^{2}$. This turns out to reasonably reproduce
the curve $nr_{d}(\alpha)$ obtained from our numerical results (not
shown here).

\subsection{Inhomogeneous case\label{sub:Cluster-inh}}

Let us now develop a model that goes beyond the intuitive criterion
(\ref{eq:cluster-criterion_Ldx}) by including the CM dependence of
the energy.

\subsubsection{Bosons \label{sub:Cluster-inh-Bosons}}

So far, in assuming a net attraction $\bar{g}<0$, we have limited
ourselves to a cluster state already tightly localized in the potential
minimum of $V_{\mathrm{CM}}$. More generally, a CM distribution smeared
out about $\Theta=\pm\frac{\pi}{2}$ will shift the average interaction
energy upward, possibly even to positive values, i.e., $E_{\mathrm{int}}\sim\frac{N^{2}}{2L}\langle g(X\equiv R\Theta)\rangle$
with \[
\langle g(X)\rangle=g(R\frac{\pi}{2})+g(0)3\sin^{2}\alpha\langle\sin^{2}(\Theta-\frac{\pi}{2})\rangle\equiv\bar{g}+\delta g.\]

For a droplet much smaller than the ring radius, $L\ll R$, we have
$\langle\sin^{2}(\Theta-\frac{\pi}{2})\rangle\simeq\frac{1}{4}\left(\frac{L}{R}\right)^{2}$.
This regime applies to a tightly bound cluster with $\alpha\gtrsim\alpha_{c}$
or $nr_{d}$ well above the critical value. Then the total energy
has an interaction term $\propto L\times N^{2}g(0)\sin^{2}\alpha/R^{2}$
corresponding to a linear confinement of the CM, in addition to the
attractive term known from the homogeneous case. For convenience,
the total energy can be cast in the form \[
\frac{E}{N\frac{\hbar^{2}}{mL^{2}}}\sim1-\frac{LN}{\bar{a}_{1}}\left(1-\frac{L^{2}}{\delta x^{2}}\right),\]
 where \[
\left(\frac{\delta x}{R}\right)^{2}\equiv-\frac{\bar{g}}{\delta g}=4\left(1-\left|\frac{\sin\alpha_{c}}{\sin\alpha}\right|^{2}\right)\]
 happens to give the leading order of the attractive-zone width $\Delta x$
for $\alpha\to\alpha_{c}$, and differs only marginally even for larger
$\alpha$. For $\bar{a}_{1}/N\ll\delta x$, that is, a broad attractive
zone which would easily accommodate a cluster with free-space binding
length $\bar{a}_{1}/N$, the solution is \[
L\simeq\frac{\bar{a}_{1}}{N}\left(1-\left|\frac{\bar{a}_{1}/N}{\delta x}\right|^{2}\right).\]
To leading order, this recovers a droplet with the free-space binding
length, but slightly compressed due to the confining term, repelling
the CM from the walls of $V_{\mathrm{CM}}$ as the CM wavepacket spreads
out beyond the potential minimum. Conversely, in the case where the
free-space cluster would have a size strongly exceeding the attractive
region, $\bar{a}_{1}/N\gg\delta x$, the droplet size is indeed much
smaller than that: \[
L\simeq\sqrt[3]{\delta x^{2}\bar{a}_{1}/N}.\]
In this limit, the total energy is clearly positive, and the object
is held together by the inhomogeneity---the CM confinement---rather
than the attractive mechanism responsible in the homogeneous case. 

Note that our above estimate required $L\ll R$, so this puts a bound
on how weakly confined the dipoles may be. For larger droplet size
$L\gtrsim R$, the CM-confinement term $\delta g$ will become weaker
until eventually, as $L\gg R$, it will saturate to $\langle\sin^{2}(\Theta-\frac{\pi}{2})\rangle\to\frac{1}{2}$.
In that limit, the minimum value $\bar{g}$ will simply be replaced
by the uniform average $\langle g\rangle=\bar{g}+g(0)\frac{3}{2}\sin^{2}\alpha=(1-\frac{3}{2}\sin^{2}\alpha)g(0)$.
This is exactly equivalent to the result obtained the homogeneous
case (\ref{eq:MFshift}): A bound state can only exist for $\langle g\rangle<0$,
i.e., in the parameter regime of average attraction, $\alpha>\alpha_{c2}$.
Its binding length is then given by the analog of (\ref{eq:binding-length}),
with $\bar{a}_{1}\mapsto-2\hbar^{2}/m\langle g\rangle$ significantly
larger due to the weaker binding. Since $L\gtrsim R$ by assumption,
this can describe a cluster state with $L<\Delta x$ at the utmost
for large $\alpha\approx\pi/2$.

\subsubsection{Fermions}

For fermions, assuming a local-density approximation, the inhomogeneity
will have a similar effect, namely, to smear out the position of the
cluster about the minimum, so that the average direct-interaction
term is increased, $\langle g(X)\rangle=\bar{g}+\delta g$. Under
the assumptions made in Sec.~\ref{sub:Cluster-hom}, this leads to
the same results as in the bosonic case but with $\bar{a}_{1}/N$
replaced by $N\bar{a}_{1}$: For $N\bar{a}_{1}\ll\delta x$, the cluster
size $L$ is slightly compressed compared with the homogeneous estimate,
$N\bar{a}_{1}$. In the opposite limit, $L\simeq\sqrt[3]{\delta x^{2}N\bar{a}_{1}}$
is somewhat larger than the attractive zone, corresponding to a repulsively
bound object. Of course, these estimates are valid only for moderate
sizes, $a_{\perp}\ll L\lesssim Na_{\perp}$, as discussed above.

\subsection{Two-mode description\label{sub:Two-mode-description}}

After having derived a simple model for the single-cluster state,
let us now take a broader perspective on the different phases we found
in the regime of strongly inhomogeneous, partially attractive interactions.
For simplicity, let us focus on bosons; the fermionic case can be
discussed analogously. 
\begin{itemize}
\item For small enough coupling, the state is always that of a weakly interacting
gas. For bosons, this is a Bose gas with net attraction for $\alpha>\alpha_{c2}$
(net repulsion otherwise), distributed coherently over the regions
$x_{\pm}=\pm\pi R/2$. If the single-particle orbital occupied by
all bosons can be written in terms of two modes localized at $x_{\pm}$,
$\varphi_{0}(x)=\frac{1}{\sqrt{2}}\left[\varphi_{+}(x)+\varphi_{-}(x)\right]$,
each created by the field operator $a_{\pm}^{\dagger}$, then the
many-body ground state takes the form \begin{equation}
|N\rangle_{0}=\frac{1}{\sqrt{N!}}\left(\frac{a_{+}^{\dagger}+a_{-}^{\dagger}}{\sqrt{2}}\right)^{N}|0\rangle.\label{eq:SF}\end{equation}

\item For sufficiently large $nr{}_{d}$ , a cluster state is formed for
any $\alpha>\alpha_{c}$. This may be written as a cat state \begin{equation}
\frac{1}{\sqrt{2}}\left(|N,0\rangle+|0,N\rangle\right)\equiv\frac{1}{\sqrt{2N!}}\left(a_{+}^{N}+a_{-}^{N}\right)^{\dagger}|0\rangle.\negmedspace\label{eq:cat}\end{equation}

\item Furthermore, for small enough $\alpha\gtrsim\alpha_{c}$, there is
a transitional regime for intermediate coupling where the ground state
has a Mott-insulator-like character: That is, half of the dipoles
are localized near $\theta=+\pi/2$ and the other half near $-\pi/2$
(taking $N\in2\mathbb{N}$ for simplicity), and the wave function
approximately has the structure \begin{equation}
|{\textstyle \frac{N}{2}},{\textstyle \frac{N}{2}}\rangle=\frac{1}{\frac{N}{2}!}\left(a_{+}^{\dagger}\right)^{N/2}\left(a_{-}^{\dagger}\right)^{N/2}|0\rangle.\label{eq:MI}\end{equation}

\end{itemize}
Expanding the many-body Hamiltonian $H$ in terms of the two-mode
states $|N_{+},N_{-}\rangle$, one obtains an effective Hubbard-like
Hamiltonian \begin{equation}
H_{J,U}=-J\sum_{\langle s,s'\rangle}a_{s}^{\dagger}a_{s'}+\frac{U}{2}\sum_{s}\hat{n}_{s}(\hat{n}_{s}-1),\label{eq:BHM}\end{equation}
 where $J\equiv\langle\varphi_{-}|$$\frac{p^{2}}{2m}|\varphi_{+}\rangle$
denotes an effective tunnel coupling, $U\equiv\langle\varphi_{s},\varphi_{s}|V(x,x')|\varphi_{s},\varphi_{s}\rangle$
is the on-site interaction, and $\hat{n}_{s}\equiv a_{s}^{\dagger}a_{s}$
$(s=\pm)$. As in the derivation of the Hubbard model in the context
of external (periodic or double-well) potentials \cite{milburn97},
this implicitly assumes tight binding, which requires both tunnel
coupling and on-site interaction to be negligible compared with the
excitation energy of higher {}``bands'' (i.e., excited states pertaining
to the doublet $\varphi_{\pm}$). Moreover, we have suppressed an
overall kinetic-energy offset $\sim N\hbar^{2}/mL^{2}$ and an off-site
interaction $U_{\pm}\hat{n}_{+}\hat{n}_{-}$, falling off as $(a_{\perp}/R)^{3}$.

It is clear from the considerations in Sec.~\ref{sub:Cluster-hom}
that the orbitals $\varphi_{\pm}$ (and thus $J,U$) are not known
a priori but rather depend on $N$ and $D^{2}$. In this sense, diagonalizing
the effective Hamiltonian $H_{J,U}$ only yields self-consistent information
on the ground state. Nonetheless, mapping the phases of the two-site
Bose-Hubbard model \cite{mueller06} to our system may provide some
useful insight:
\begin{itemize}
\item For weak on-site interaction, $N|U|\ll J$, the ground state is a
delocalized superfluid state (\ref{eq:SF}), maximizing the coherence
$\langle a_{+}^{\dagger}a_{-}\rangle$. 
\item In the limit of strong attraction, $|U|\gg J$, the ground state localizes
on a single site, thus maximizing the absolute interaction energy
$\sim|U|N^{2}/2$. By symmetry, the state will be a superposition
of all single-site states, corresponding to the cluster state (\ref{eq:cat}).
\item For strong repulsion, $NU\gg J$, the energy is dominated by the on-site
repulsion. That is minimized by a fragmented or Mott-insulator state
(\ref{eq:MI}) with interaction energy $\sim U(N/2)^{2}$.
\end{itemize}
This indicates that the transition from a fragmented to a clustered
state should take place when $U\sim\langle g(X)\rangle/L$ switches
from positive to negative. Clearly, this cannot occur for $\alpha>\alpha_{c2}$,
in which case there is a direct crossover from gas-like to clustered
behavior. Near the critical angle in turn, $\alpha_{c}\lesssim\alpha<\alpha_{c2}$,
$U$ changes from positive to negative at $\bar{a}_{1}/N\lesssim\Delta x$,
i.e., for $nr_{d}$ given by the criterion (\ref{eq:cluster-criterion_nrd}).
Below that critical coupling, the {}``repulsively bound'' cluster
state (see Sec.~\ref{sub:Cluster-inh-Bosons}) is an excitation on
top of the insulator-type state, whereas for higher $nr_{d}$, the
attractively bound cluster is the ground state.

\section{Conclusion and outlook}

In this article, we have shown that dipolar particles confined to
a quasi-1D ring exhibit inhomogeneous gas-, solid- and (clustered)
droplet-like ground states, due to the underlying anisotropy of the
3D dipole interaction. 

The solid-like behavior is essentially independent of the transverse
confinement, but differs from that in a linear geometry by a non-equidistant
lattice-site distribution whenever the dipoles are inclined toward
the plane of the ring ($\alpha\neq0$). By contrast, especially the
Bose-gas-like states depend crucially on the fact that the transverse
confinement length cuts off the \emph{$|\mathbf{r}|^{-3}$} divergence
of the dipole interaction. This leads to a whole class of states with
a tunable interaction parameter, as described by the (inhomogeneous)
Lieb-Liniger model, rather than merely a hard-core gas. By the same
token, the appearance of attractive interaction regions for dipoles
oriented sufficiently close to the plane of the ring leads to bound
states clustered around the {}``poles'' of the ring, rather than
collapse. 

Our numerical results are presented for few particles; however, complementary
analytical models support an extrapolation to many-body systems. All
results are based on a Born approximation for the effective 1D interaction,
i.e., on the occupation of a single transverse mode. An interesting
subject of future investigation may thus be how an interaction-induced
coupling of several transverse modes affects these results. Moreover,
our calculations provide motivation for experimental studies as well
as theoretical extensions to dipolar systems in more general curved
geometries (including, e.g., coupled quasi-1D systems \cite{wunsch11,dalmonte11}).

\begin{acknowledgments}
The author is indebted to C.~J. Pethick for many inspiring dicussions
and comments on the manuscript. Special thanks also go to M.~Girardeau,
A.~Griesmaier, H.-D.~Meyer, and N.~Zinner. Financial support from
the German Academy of Sciences Leopoldina (LPDS 2009-11) is gratefully
acknowledged.
\end{acknowledgments}
\appendix

\section{Computational method\label{sec:method}}

We use the numerically exact multi-configurational time-dependent
Hartree method \cite{bec00:1}, a quantum-dynamics tool which has
been applied to few-body systems of identical bosons as well as mixtures
(see, e.g., \cite{zoellner06a} for details). Its principal idea is
to solve the time-dependent Schrödinger equation $\begin{array}{c}
i\hbar\dot{\Psi}(t)=H\Psi(t)\end{array}$ as an initial-value problem by expanding the solution in terms of
direct (or Hartree) products $\Phi_{\boldsymbol{j}}\equiv\varphi_{j_{1}}\otimes\cdots\otimes\varphi_{j_{N}}$:\begin{equation}
\Psi(\boldsymbol{\theta};t)=\sum_{\boldsymbol{j}}A_{\boldsymbol{j}}(t)\Phi_{\boldsymbol{j}}(\boldsymbol{\theta};t).\label{eq:mctdh-ansatz}\end{equation}
Both coefficients $A_{\boldsymbol{j}}(t)$ and the single-particles
basis functions $\varphi_{j}(t)$ are determined from the Dirac-Frenkel
variational principle $\langle\delta\Psi|[i\hbar\partial_{t}-H(t)]\Psi(t)\rangle=0$
\cite{bec00:1}. This leads to a coupled system of Schrödinger- and
mean-field-type equations for $A_{\boldsymbol{j}}$ and $\varphi_{j}$,
respectively, which is integrated numerically upon discretization.
Note that the expansion coefficients inherit the (bosonic or fermionic)
permutation symmetry of the wave function, $A_{P(\boldsymbol{j})}=(\pm1)^{\mathrm{inv}(P)}A_{\boldsymbol{j}}.$
Moreover, in our case of particles on a ring, the single-particle
wave functions $\varphi_{j}\in\mathrm{L}^{2}(-\pi,\pi)$ are periodic
under translations $\theta\mapsto\theta+2\pi$, which is ensured by
expanding them in an appropriate basis set.

Although designed for time-dependent simulations, it is also possible
to apply this approach to stationary states. This is done via \emph{relaxation},
i.e., by applying the non-unitary imaginary-time evolution operator
$e^{-H\tau}$. As $\tau\to\infty$, this exponentially damps out all
contributions as $e^{-(E_{m}-E_{0})\tau/\hbar}$ except that from
the ground state. In practice, one relies on a more robust scheme
termed \emph{improved relaxation} \cite{mey03:251}. Here $\langle\Psi|H|\Psi\rangle$
is minimized with respect to both the coefficients $A_{\boldsymbol{j}}$
and the orbitals $\varphi_{j}$. The effective eigenvalue problems
thus obtained are then solved iteratively by first solving for $A_{\boldsymbol{j}}$
with \emph{fixed} orbitals and then {}`optimizing' $\varphi_{j}$
by propagating them in imaginary time over a short period. That cycle
is then repeated.

\bibliographystyle{../../../bin/prsty}
\bibliography{../../../bib/phd,../../../bib/mctdh}

\end{document}